\documentclass[format=acmsmall,screen,review=false,nonacm=true]{acmart}
\AtBeginDocument{%
  \providecommand\BibTeX{{%
    \normalfont B\kern-0.5em{\scshape i\kern-0.25em b}\kern-0.8em\TeX}}}

\setcopyright{none}
\copyrightyear{}
\acmYear{}
\acmDOI{}

\acmPrice{}
\acmISBN{}

\usepackage{caption}
\usepackage{subcaption}

\usepackage{lscape}

\begin{document}

\title{Designing for Sustained Motivation: A Review of Self-Determination Theory in Behaviour Change Technologies}

\author{Lize Alberts}
\email{lize.alberts@cs.ox.ac.uk}
\affiliation{%
  \institution{Department of Computer Science, University of Oxford}
  \city{Oxford}
  \country{UK}
}
\affiliation{%
  \institution{Department of Philosophy, Stellenbosch University}
  \city{Stellenbosch}
  \country{ZA}
}
\author{Ulrik Lyngs}
\email{ulrik.lyngs@cs.ox.ac.uk}
\affiliation{%
  \institution{Department of Computer Science, University of Oxford}
  \city{Oxford}
  \country{UK}
}
\author{Kai Lukoff}
\email{klukoff@scu.edu}
\affiliation{%
  \institution{Department of Computer Science and Engineering, Santa Clara University}
  \city{Santa Clara}
  \state{California}
  \country{USA}
}


\begin{abstract}

Recent years have seen a surge in applications and technologies aimed at motivating users to achieve personal goals and improve their wellbeing. However, these often fail to promote long-term behaviour change, and sometimes even backfire. We consider how self-determination theory (SDT), a metatheory of human motivation and wellbeing, can help explain why such technologies fail, and how they may better help users internalise the motivation behind their goals and make enduring changes in their behaviour. In this work, we systematically reviewed 15 papers in the ACM Digital Library that apply SDT to the design of behaviour change technologies (BCTs).
We identified 50 suggestions for design features in BCTs, grounded in SDT, that researchers have applied to enhance user motivation. However, we find that SDT is often leveraged to optimise engagement with the technology itself rather than with the targeted behaviour change \textit{per se}. When interpreted through the lens of SDT, the implication is that BCTs may fail to cultivate sustained changes in behaviour, as users' motivation depends on their enjoyment of the intervention, which may wane over time. An underexplored opportunity remains for designers to leverage SDT to support users to internalise the ultimate goals and value of certain behaviour changes, enhancing their motivation to sustain these changes in the long term.
\end{abstract}



\maketitle

\section{Introduction}

In recent years, there has been a surge in applications (e.g., Duolingo, Calm, Forest), browser extensions (e.g., Newsfeed Eradicator), and other technologies (e.g., wearables like FitBit) that are designed to help users bring about positive behavioural changes in their lives, including losing weight, gaining new skills, quitting addictions, or developing mindfulness. These are sometimes referred to under the broader class of \textit{persuasive technologies} \cite{sundar2012motivational}, or the narrower \textit{behaviour change apps} \cite{Villalobos} or \textit{mobile health (mHealth) technologies} \cite{Molina}. In this paper, we use the term \textit{behaviour change technologies} (BCTs) from Hekler \textit{et al.} \cite{Hekler} to refer to ``the broad array of systems and artifacts developed to foster and assist behavior change and sustainment'' \cite[p.3308]{Hekler}. 
These technologies employ a range of techniques and features (e.g. reminders, rewards, performance tracking) to motivate, persuade, nudge and otherwise support users in performing or avoiding certain behaviours. 
 
As of 2021, there are around 350k mHealth (including medical, fitness, and mental health) apps hosted across different app stores \cite{report}, and over 500M people use these to monitor and/or regulate their daily activities \cite{Villalobos}. Of those, around 55K apps are currently available on Google Play and 41K on the Apple App Store (with numbers peaking around the pandemic) \cite{statista}.  Other major categories of BCTs include education apps that motivate learning (which make up the third biggest sector in the Apple App Store \cite{app}), and digital self-control tools (DSCTs) like apps or browser extensions that help millions of people limit unwanted time spent on their devices or online \cite{Lyngs2019SelfControlInCyberspace}. 

However, despite their popularity, BCTs often fail to help individuals bring about the intended positive changes in the long term \cite{mustafa2022user,harrison2015activity}. 
For example, a recent study found that about 53\% of mHealth apps were uninstalled within 30 days of download \cite{mustafa2022user}. 
In surveys, aside from lack of desired features, boredom and loss of motivation are often reported as reasons for behaviour change app abandonment \cite{honary2019understanding}. 
This suggests that initial use is driven by a novelty effect, but that these technologies lack reliable ways to motivate sustained use \cite{mustafa2022user,schmidt2019gamification,tsay2020overcoming}.

Another factor is the severity of enforcement or degree of friction that a BCT uses to hold users accountable to their goal  \cite{Cox2016DesignFrictions, Kim2019Goalkeeper}: too weak and it might be too easy for the user to circumvent the intervention (e.g., ignoring a reminder sent by the app), too strong and it might trigger frustration and lead them to abandon the tool completely (e.g., blocking a user's smartphone after a daily usage limit has been hit, with no override option) \cite{Lukoff2022-xs}. 
Thus, work on BCTs for helping people reduce time on distracting websites has found that people typically begin by setting themselves highly challenging interventions, but often reject enforcement at the moment of temptation and slip into easier interventions \cite{Kovacs2021NotNowAskAgainLater}. 
 
In some cases, a BCT can even backfire, causing the opposite behaviour to what was intended (e.g. \cite{Lyngs2020FacebookHack,erskine2010suppress,stibe2016persuasive}).\footnote{For example, in a study of goal reminders for supporting self-regulated Facebook use, a participant said the intrusiveness of the intervention made her want to ``stay on just out of spite'' \cite{Lyngs2020FacebookHack}.} Such responses could be understood as manifestations of psychological \textit{reactance} \cite{Fitzsimons2004-qx,McAlaney2020-dc}: unpleasant motivational arousal to situations that threaten certain behavioural freedoms. This reactance\footnote{A.k.a the “boomerang effect,” or more colloquially, the “screw you, don’t tell me what to do” effect \cite{Isaac2021-ky}.} may result in ``behavioural backlash,'' when a person not only fails to comply with expectations, but intentionally contradicts them \cite{Fitzsimons2004-qx}. 

Another contributing factor may be the goals and metrics that designers use to evaluate the success of BCTs. For designers, a primary concern tends to be maximising engagement with the app, which is not only a straightforward way to measure success, but is typically also in the interest of the app developer. In a recent survey of BCTs on the Apple App Store, Villalobos-Z\'{u}\~{n}iga and Cherubini \cite{villa} found that reminders are the most popular feature apps use to motivate people to keep to their goals. They maintain that it is likely overused for specifically that reason: to keep users returning to the app. As such, more effort may be spent on maximising engagement with the technology than on facilitating sustained behaviour change. 

This raises some important questions for the design of BCTs, namely: \textit{How can BCTs support sustained behaviour change? How can BCTs help regulate behaviour without making users feel their autonomy is threatened? What metrics of success can be used to evaluate user benefit?} In this paper, we explore how self-determination theory (SDT), an evidence-based theory of human motivation, growth, and wellbeing \cite{deci1985intrinsic, ryan2000intrinsic,ryan2000self}, can be leveraged to help answer such questions. 

SDT has emerged as among the most researched and applied theoretical framework of human motivation in psychology today \cite{ryan2022we}. It has been actively and increasingly examined for over four decades, with many of its central tenets repeatedly evaluated in meta-analyses \cite{ryan2022we,metagillison,metantoumanis}. This strong interest in the theory is partly attributable to its relatively unique focus on the importance of human autonomy and volition: how environmental factors can either support or undermine a person's ability to willingly and proactively pursue a behaviour in an enduring way \cite{ryan2022we}. 

SDT proposes specific factors that affect people's ability to develop and sustain motivation for specific behaviours, decreasing their dependence on support from external sources. On a basic level, the theory maintains that sources of motivation can be more or less internal (coming from the individual's inherent desires to perform the activity) or external (coming from outside the individual, like fears or rewards or senses of obligation). 
According to SDT, whilst interventions based on extrinsic forms of motivation may work in the short term, they typically fail to sustain motivation in the long term, or after an intervention has ended \cite{deci1985intrinsic}. Conversely, when motivation becomes internalised, actions can become self-determined, at which point interventions may no longer be needed, and changes in behaviour tend to persist \cite{ryan2017self}. A major opportunity for BCTs is to leverage SDT to change the nature of the scaffolding mechanisms towards more internal regulation of motivation. That is, to help people internalise the values and goals of the intervention such that they feel less like it is something they \textit{have to} do, but rather, something they do voluntarily as they find it intrinsically enjoyable and/or personally meaningful. According to subtheories within SDT, this change is facilitated by the relevant satisfaction of three \textit{basic psychological needs} (BPNs) that all humans seem to have: needs for a sense of autonomy, competence, and relatedness. These need satisfactions provide the essential nutrients for energising more autonomous forms of regulation, which SDT posits as necessary mechanisms that underlie long-term changes in behaviour \cite{ryan2017self,metagillison}. By setting out these mechanisms, SDT provides a framework for developing behaviour interventions, and proposes techniques for influencing its constructs \cite{Fortier_Duda_Guerin_Teixeira_2012,metagillison,metantoumanis,ryan2022we}.

There is strong evidence for the efficacy of SDT-based interventions across various domains.\footnote{This includes physical activity \cite{edmunds2008testing,wilson2006formative}, environmental behaviours \cite{pelletier2008persuasive}, tobacco addiction \cite{williams2009importance}, and healthcare treatment adherence \cite{williams2004testing}---see \cite{metagillison,metantoumanis,ryan2022we} for meta-analyses of SDT-based intervention studies.} As such, HCI researchers have started to highlight the utility of incorporating SDT constructs in interface design, and what it may mean to support these basic needs at different stages of technology adoption \cite{Ballou2022SDTinHCI,Peters2018-ra}. For BCTs in particular, a handful of HCI papers have started to consider how these needs may be supported with design features. To evaluate how this has been done and what gaps remain for future work, particularly towards supporting sustained behaviour change, we surveyed publications in flagship HCI venues with the following research questions:



\begin{quote}
\textbf{(RQ1) }Which of SDT's theories have HCI researchers applied to the design of behaviour change technologies? \\
\textbf{(RQ2) }What reasons were given for the application of SDT?\\
\textbf{(RQ3) }How have SDT-related constructs been translated into design features?\\
\end{quote}

For this, we systematically reviewed all papers in the ACM Digital Library that mention SDT in the abstract, including in our analysis any that apply the theory towards the design of technologies that have the purpose of helping individuals change their behaviour in some personally desired way.  We drew inspiration from Tyack and Mekler's \cite{games} recent review of how SDT has been applied in HCI in the context of gameplay, which highlights conceptual gaps and limitations, as well as opportunities for applying the theory in that space. 

Given the breadth of application domains and aims of BCTs, we also consider a fourth question:
\begin{quote}
\textbf{(RQ4) }What contextual factors need to be considered to evaluate whether these design features are suitable for different BCTs?  
\end{quote}
Across the 15 papers we reviewed, we identified 50 design suggestions: 11 for supporting `autonomy', 22 for `competence', and 17 for `relatedness'. 
 We also identified some common inconsistencies between the broader aims and purpose of the constructs, as defined by SDT, and how it was translated into design. 
 Despite SDT’s central aim of supporting integrated forms of motivation, we found that BCTs tend to utilise it more for the sake of making the \textit{technology} as inherently engaging and satisfying as possible, than for offering scaffolding, information and support for helping users internalise the target behaviour \textit{per se}. Whilst most papers consider how to support the basic psychological needs that SDT postulates (often through gamification tactics that make \textit{doing} the tasks more enjoyable), few consider how to help users internalise the deeper value of the behaviour change and bring it into congruence with their other goals and values, such that they are intrinsically motivated to keep pursuing the target behaviours when the intervention ends. 
 
 By fostering further reliance on extrinsic forms of motivation, these technologies are only able to sustain behaviour changes as long as users find the technology enjoyable and satisfying, which holds several risks. Firstly, if, for some reason, the users are unwilling or unable to keep using the BCT (e.g., as it gets discontinued or becomes too expensive), they may no longer find enjoyment in pursuing the target behaviour. Secondly, if a BCT only makes the target behaviour enjoyable (e.g., with a satisfying interface design) without helping users understand and appreciate its ultimate value, their enjoyment may wane or be replaced by something else they find more enjoyable (and requires less effort). Moreover, users may be incentivised to perform the behaviours for the wrong reasons, perhaps undermining the ultimate goal (e.g., hacking a step-counter to seem more active, or exercising to the detriment of their overall health to keep their streak). In many cases, it may be more desirable for people to internalise their motivation for meeting their personal goals and knowledge of how to achieve them, such that they no longer depend on the BCT. 

 By reviewing papers in the domain of HCI, we were able to understand how researchers have applied SDT to the design of technological interventions. By contrast, when SDT is leveraged by papers in other fields (e.g., health), outcomes are usually reported, but the BCT itself is often described in insufficient detail to allow for an analysis of design considerations (RQs 3 and 4). One risk we identified is that HCI researchers in this space sometimes employ design features that relate to the theory by association (e.g., understanding relatedness as any form of social interaction), without using measures to ensure that the features actually support the desired construct effectively. This is a common pitfall when applying behavioural theories to technology design more broadly \cite{michie2010interventions,Hekler,michie,roff}, and has also been observed in meta-analyses of SDT-based interventions in other domains \cite{metagillison}. This may keep research interventions from achieving the sustained motivation that the theory promises. Drawing from Hekler et al. \cite{Hekler} and others, we offer suggestions for research practices to help ensure that the theory is applied most usefully. 

Whereas recent work suggests ways to integrate assessment of SDT's autonomy construct when designing wellbeing-supportive technologies \cite{Peters2018-ra}, we consider implications for supporting user motivation and wellbeing in the special case of BCTs, where a technology is purposefully designed/used to support certain desired behavioural changes, and, preferably, in ways that can be sustained. Our paper also contributes to the growing body of research in the design and evaluation of BCTs and for HCI researchers interested in utilising SDT.

In particular, we contribute:
\begin{enumerate}
    \item   A systematic review of how SDT has been applied to the design of behaviour change technologies in HCI, especially regarding claims about supporting core SDT constructs and facilitating sustained motivation by design;
    \item An overview of contextual factors that may influence the suitability of SDT-inspired design features;
    \item  Recommendations for how the HCI community might unlock more of the theory's potential, drawing from Hekler \textit{et al.}'s suggestions for using behavioural theory in HCI.
\end{enumerate}

\section{Self-determination theory}

SDT is proposed as an `organismic' metatheory of human motivation.\footnote{Motivation, here, is understood as being \textit{moved} (inspired, energised) to act \cite{ryan2000intrinsic}.} That is, it describes motivation in terms of its relation to the evolved (human) organism as a whole, rather than treating it as an isolated phenomenon: 
\begin{quote}
In essence, SDT attempts to articulate the basic, vital nature of human beings— of ­ how that nature expresses itself, what is required to sustain energy and motivation, and how that vital energy is depleted \cite[p.24]{ryan2017self}.
\end{quote}
It is based on the observation that people have an innate propensity to be proactive, creative and self-motivated---striving to grow, learn, master new skills, and express their talents \cite{ryan2000self}. 
This natural growth-tendency is understood in terms of our more general evolved resources for personality development and behavioural self-regulation. 
However, this propensity for self-motivation seems to be sustained only under certain social and environmental conditions. Under others, people can be passive and demotivated, rejecting growth \cite{ryan2000self,ryan2000intrinsic}. 
The importance of these supporting conditions bears on the particular view of motivation the theory posits. 
More than just varying in \textit{degree} (how much motivation), SDT describes different \textit{types} of motivation, pertaining to differences in the ``why'', i.e., the underlying attitudes and reasons that move people to act \cite{ryan2000intrinsic}. Within the six sub-theories of SDT, the BPNs play a key mediating role between different motivation types.

\subsection{Types of motivation}
The most basic distinction SDT makes is between \textit{intrinsic motivation}, ``doing something because it is inherently interesting or enjoyable'' \cite[p.55]{ryan2000intrinsic}, and \textit{extrinsic motivation}, ``doing something because it leads to a separable outcome''  \cite[p.55]{ryan2000intrinsic}. For instance, a child could practice playing the violin (mainly) because they find it inherently fun and satisfying to improve their skills, or because of a range of external aims or pressures (e.g., because their parents force them to, because they want to impress their friends, because they are afraid of the violin teacher shouting at them, because they want to achieve a good grade). 

Variability in intrinsic motivation is described by one of SDT's subtheories, Cognitive Evaluation Theory (CET). According to CET, when someone is intrinsically motivated to apply themselves in some way, they tend to be at their most creative, energised and driven. In such cases, not only is the quality of outcome typically higher, but their motivation for the behaviour will persist for longer. 
In contrast, when someone is purely motivated by extrinsic factors, they may apply themselves with a sense of ambivalence, boredom and even resentment \cite{ryan2000intrinsic}. In its most general form, CET maintains that the fundamental needs for competence and autonomy are key factors that distinguish intrinsic and extrinsic motivation: feeling capable of doing the task, and feeling like one is willingly doing it for personal reasons. The theory further maintains that both competence and autonomy satisfactions are necessary for sustaining intrinsic motivation. In addition, in observing that ``intrinsic motivation is most robust in a context of relational security and can be enhanced by a sense of belonging and connection'' \cite[p.124]{ryan2017self}, CET suggests that relatedness also plays a role in intrinsic motivation, especially for activities that involve a social element.\footnote{Several evaluations have reported a comparatively small but non-significant effect of relatedness satisfaction \cite{metagillison,metantoumanis}. Ntoumanis \textit{et al.} \cite{metantoumanis} suggest that, whilst relatedness supports ``might be useful to support initiation of change but perhaps not maintain it long-term, particularly if the target behaviour is complex or does not need to take place alongside other people (e.g., being regularly physically active, eat healthy)'' \cite[p.234]{metantoumanis}. In other domains (e.g., social contexts like work environments) meta-analytic evidence suggests a more significant effect of relatedness support \cite{metantoumanis}. 
}

Whilst desirable, an individual will not be intrinsically motivated to perform just any activity. Ryan and Deci maintain that intrinsic motivation has a lot to do with the sort of personality and values an individual has: it ``exists in the relation between individuals and activities"  \cite[p.56]{ryan2000intrinsic}, as individuals may naturally care more easily about certain tasks than others (e.g., learning a new skill versus doing their taxes). In cases where some external sources of motivation are required (be it encouragement, deadlines, or fines), people may still be supported in ways that foster a greater or lesser sense of willingness and drive. 
Another subtheory of SDT, Organismic Integration Theory (OIT), describes four types of motivation within the broader category of extrinsic motivation \cite{ryan2000self}. These types differ in their degree of \textit{internalisation}---the degree to which the behavioural regulation is experienced as autonomous versus controlled. The key determining factor here is what is called the \textit{perceived locus of causality}, i.e., where the source of regulation is perceived to be (i.e., coming mainly from outside or within).

The least autonomous form of extrinsic motivation is called \textit{external regulation} (doing something purely ``to satisfy an external demand or obtain an externally imposed reward contingency'' \cite[p.61]{ryan2000intrinsic}. 
A slightly more autonomous type is \textit{introjected  regulation}, when an activity is done out of a sense of pressure, typically to ``avoid guilt or anxiety, or attain ego-enhancements or pride''  \cite[p.62]{ryan2000intrinsic}. 
According to OIT, both of these types still have an \textit{external perceived locus of causality} as the source of motivation is mainly external pressures or incentives. 

The remaining two types have more of an \textit{internal perceived locus of causality}. 
The first is \textit{identification}, i.e., when a person identifies with the personal importance of a behaviour such that they accept the regulation to come from within \cite{ryan2000intrinsic}. 
This happens when the outcome of an activity is required for something that the person deems personally important, e.g., filling out an application form for the sake of a job they really want. 
The most autonomous type is \textit{integrated regulation}, which happens when, through a process of self-examination and bringing regulation's aims in congruence with their other needs/values, the regulation has been ``fully assimilated into the self'' \cite[p.62]{ryan2000intrinsic}. 
Like intrinsic motivation, integrated forms of motivation lead to behaviours that are volitional and unconflicted, although they differ from intrinsic motivation in that they remain valued for their \textit{instrumental} value to serve some separate goal, rather than pursuing the activity for its \textit{own} inherent value \cite{ryan2017self}. 

\begin{figure}
    \centering
    \includegraphics[scale=0.45]{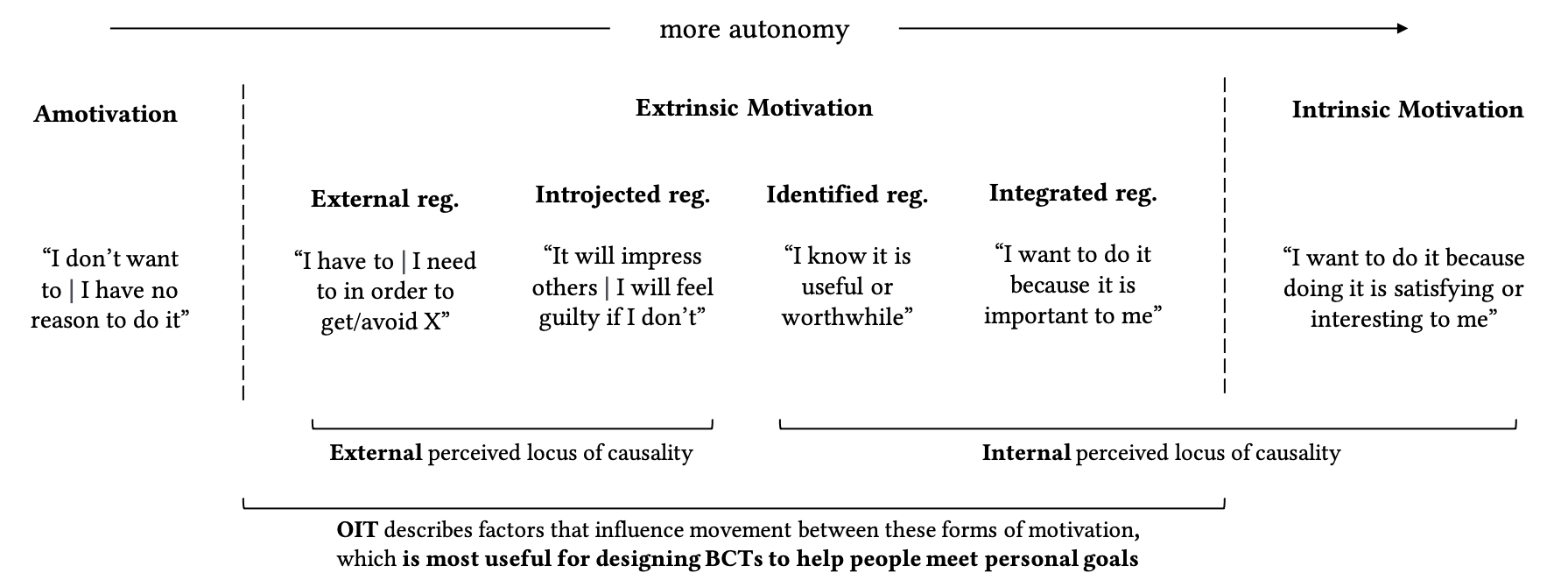}\\
    \caption{Motivation types postulated by SDT, ranging from amotivation (least autonomous) to intrinsic motivation (most autonomous). Adapted from \cite{Peters2018-ra} and \cite{ryan2000self}.}
    \label{fig:motivation}
\end{figure}

These four types of \textit{extrinsic motivation} fall on a continuum between \textit{amotivation} (lacking any intention to act/sense of control) and \textit{intrinsic motivation} (as defined above)---see Fig. \ref{fig:motivation}.  

According to OIT, the quality and strength of motivation can increase or decrease depending on the extent to which a person's relative need for \textit{autonomy} is supported by relevant environmental conditions: it is the key mediating factor between the different extrinsic motivation types \cite{ryan2000intrinsic,ryan2017self}. 
As for the other BPNs, they further support the process of internalisation in different ways. One is by offering justifications for the importance of valuing a certain behaviour, e.g., understanding why it is personally meaningful (\textit{competence}) or that it can bring them closer to people want to be connected with (\textit{relatedness}) \cite{ryan2000intrinsic,ryan2017self}. Another is by feeling efficacious concerning it: as if one has the relevant skills to succeed at it, and it is just the right degree of challenging (\textit{competence}) \cite{ryan2000intrinsic,ryan2000self,ryan2017self}. In short, ``[t]o fully internalize a regulation, and thus to become autonomous with respect to it, people must inwardly grasp its meaning and worth'' \cite[p.64]{ryan2000intrinsic}, which is generally facilitated by supporting someone's needs for competence and relatedness with situational factors \cite{ryan2017self}. 

However, these factors must be relevant with respect to the given domain or behaviour, and need satisfactions are implicated differently in each of the types of regulation within the OIT taxonomy \cite{ryan2017self}. For integration, what is more important than pure need support is facilitating an internal PLOC towards autonomous, integrated regulation, which relevant need-supports can help to energise and facilitate \cite{ryan2017self}. In their recent meta-analysis, Ntoumanis \textit{et al.} \cite{metantoumanis} found that changes in autonomous motivation and \textit{perceptions} of need support were associated with positive changes in health behaviours, both at the end of the intervention and at follow-up, but that changes in need-satisfaction \textit{per se} were not associated with changes in health behaviours at either time point. Following Vallerand \cite{vallerand1997toward}, they suggest that autonomous motivation may mediate the effects of psychological need satisfaction on health behaviours and people's ability to sustain them \cite{metantoumanis}. 


\subsection{Relevance for Behaviour Change Technologies}

As a theory of human motivation---the different types and how to best support it with relevant environmental conditions---SDT can help BCT designers create an environment that enhances the quality of user motivation and their ability to sustain behaviour change. 

In particular, the mini-theory of OIT is a promising theoretical framework to apply in this context. Whereas CET is useful for describing the factors that typically contribute to intrinsic motivation, OIT is most useful for understanding how to support someone to move from more external to internal sources of motivation---especially for behaviours that are required for goals they wish to achieve, but they tend to find taxing, unenjoyable or uninteresting (e.g., maintaining a healthy lifestyle, quitting addictions, etc.). These are typical reasons why people may choose to adopt BCTs.

It is also worth noting that personal goals often contain elements of both intrinsic and extrinsic motivation: whilst someone may find \textit{playing} a sport inherently satisfying, this does not necessarily apply to all aspects of what reaching their sport-related goals require (e.g., following a rigorous training regiment, managing their diet, etc.). Similarly, someone may be intrinsically motivated to play the violin in an orchestra, but not always to dedicate hours of their week to practice when they could be doing something they find more enjoyable. OIT is a useful resource for enhancing people's motivation for those aspects of meeting personal goals (as separable outcomes of behaviour change \cite{ryan2017self}) that they have the most difficulty keeping to, even if they may find it more or less intrinsically motivating at different times. As Ryan and Deci explain:
\begin{quote}
[With intrinsic motivation,] the ``aim'' is the spontaneous satisfaction experienced while doing the activity. Thus the focus is on present experience rather than future goals ... With internalized regulation, however, the focus is more on future goals or outcomes, for a defining element of extrinsic motivation is its instrumental nature, regardless of how autonomous one has become with respect to it \cite[p.197-198]{ryan2017self}.
\end{quote}
This makes OIT potentially very valuable for BCT design, as people typically adopt a BCT to meet an ultimate separable goal (e.g., losing weight, quitting smoking, gaining and maintaining a skill), rather than just finding something new and fun to do. Whereas CET may help designers understand how to make tasks more intrinsically satisfying and motivating, this is not possible for anyone for \textit{any }type of behaviour, and does not necessarily help with doing the behaviours at the desired specific and persistent frequency to meet a personal goal (i.e., the \textit{why} of behaviour change). The acknowledgement and internalisation of this \textit{why}---the ultimate value of pursuing the activity---plays a key role in facilitating integration to more autonomous regulation \cite{ryan2017self,davis2016motivating},\footnote{According to Ryan and Deci, integrated regulation ``requires self-reflection and reciprocal assimilation'' \cite[p.188]{ryan2000intrinsic}, which involves conscious endorsement of the ultimate value of the behaviour, so as to `` bring a value or regulation into congruence with the other aspects of one’s self'' \cite[p.188]{ryan2000intrinsic}.} and offers a particular design opportunity. Our work aimed to review how OIT may have been leveraged for this purpose in BCT design, or whether any of SDT's other mini-theories or constructs have been applied, and why.

One of the main challenges for BCTs is supporting long-lasting results, for which SDT is particularly well-suited.
A recent review of the main theories and models applied in HCI research on behaviour change was focused on this challenge, and highlighted the potential for, in particular, Dual Process theory, but made no reference to SDT \cite{Pinder}. 
On this background, our aim with our review is to provide a useful step towards exploring the potential of SDT in HCI research to support long-term behaviour change in service of personal user goals. More than motivation towards specific outcomes, however, the real strength of SDT lies in the fact that it treats a person as a whole organism, with particular drives, abilities, needs and goals, and who can be energised to self-sustain their motivation for activities they find personally meaningful and thrive under the right conditions. 
This supports user-centred design initiatives attempting to understand the user as a whole person, with needs and interests beyond their interaction with a given technology \cite{spiel2018fitter}, and whose overall well-being matters.

\subsection{Measures}

SDT researchers have developed numerous scales for assessing key constructs of the theory, including intrinsic and extrinsic motivation, and the three basic psychological needs. 
For example, scales that measure internalisation of extrinsic motivation typically include items that distinguish externally controlled motivation (e.g. `I do X because I might get a reward') from more autonomous motivation (e.g. `I do X because it’s important to me').
Scales have been developed for general use (e.g., the Basic Psychological Needs Satisfaction questionnaire, \cite{Chen2014BPNS}) as well as within specific domains such as work, exercise or education \cite{centerSDTQuestionnaires}.

Adaptations for HCI already exist, such as the \textit{User Motivation Inventory} (which, in the context of engagement with interactive systems, measures intrinsic motivation, integrated, identified, introjected, and external regulation, as well as amotivation, \cite{Brhlmann2018UMI}), and the \textit{Gaming Motivation Scale} \cite{Lafrenire2012GAMS}.
Measurement of autonomy is likely to be especially important for HCI research on BCTs, to avoid backfire effects from well-intentioned intervention. 
For this purpose, \cite{Peters2018-ra} have suggested a number of scale adaptations for spheres ranging from initial technology adoption (e.g., the extent to which a user’s decision to adopt a technology is volitional and personally-endorsed) 
 to larger life context (e.g., how using a technology affects a user's ability to pursue other meaningful activities in their life).

\section{Review method}
Our review aimed to investigate (\textbf{RQ1}) which of SDT's theories have been applied in HCI towards BCT design, (\textbf{RQ2}) what reasons were cited for this application, (\textbf{RQ3}) how SDT-related constructs have been translated into specific design features/suggestions, and (\textbf{RQ4}) how the suitability of these design features differs across situations. Our reviewing procedure was loosely based on the approach taken by Tyack and Mekler \cite{games}. 

\subsection{Source Selection}
Publications considered for review were drawn from the Association for Computing Machinery (ACM) Digital Library (the ACM Full-Text Collection), in order to access a broad range of key venues in HCI.\footnote{The ACM Digital Library (\url{https://dl.acm.org/}) is the most comprehensive database of computing and information technology articles and literature, covering more than 50 peer-reviewed journals in dozens of computing disciplines.} As mentioned earlier, we focused on HCI venues to get a sense of how HCI researchers have been applying SDT theory in a technological domain, to evaluate the precedents being set for interpreting the theory as design features. Whilst there may have been other relevant work in BCTs in other venues, like psychological journals, HCI has the greatest explicit focus on design considerations (RQs 3 and 4). Publications from all years were considered.

\subsection{Search procedure}
We iteratively tested the phrasing of our query to find papers relevant to our search. Initially, we searched for all papers containing the term ``self-determination'' anywhere in the text, which returned 736 results. After going through the first 146 abstracts, we found that most of the papers mentioned SDT somewhere in the text without it being central to the aims of the paper. As we were only interested in papers where SDT is explicitly applied as a lens or guiding theory, we updated our query to only include publications that contain ``self-determination theory'' in the abstract, to ensure the theory is of central importance. This returned 86 results. 

We acknowledge that this may have excluded papers where SDT was used but referred to by other means (e.g. `a popular theory of human motivation') but considered this sufficiently unlikely.   

\subsection{Screening criteria}
We did not use any exclusion criteria in our screening filters. However, we manually excluded records that merely described plans for future research. Thereby, we excluded four posters, two work-in-progress papers, two workshop proposals, one doctoral consortium paper, and one magazine article, leaving 76 records.  

\subsection{Selection of papers for inclusion in the review}
As we were interested in understanding how SDT has been \textit{applied} towards the \textit{design} of BCTs, papers had to satisfy both these conditions: (a) having substantial relevance to BCTs, (b) using SDT as a lens or guiding theory for actual system design or for suggesting design implications. 

To satisfy (a), papers had to pertain to \textit{technologies} that have the purpose of helping individuals \textit{change their behaviour} in some way that they deem \textit{personally important}. For example, we excluded papers where technologies were meant to help motivate people towards the interest of another party (e.g. motivating people for crowd work \cite{20,26,54}). Moreover, these technologies had to be ones that individuals are meant to \textit{wilfully adopt} for personal improvement---e.g., we excluded papers pertaining to technologies that are intended to be used by teachers or leaders to help a group reach outcomes, as it was not individually sought \cite{7,16,60}. However, we decided to include papers where an individual's goals are represented by someone close to them (e.g., a BCT adopted by a parent to help their young child), if the other criteria were satisfied. 

To satisfy (b), SDT (or any of its mini-theories) had to be applied specifically towards the design of a technological intervention, even just as theoretical suggestions. For example, we excluded papers where SDT was only mentioned but not applied (e.g., \cite{36,34}).

\begin{figure}
    \centering
    \includegraphics[scale=0.65]{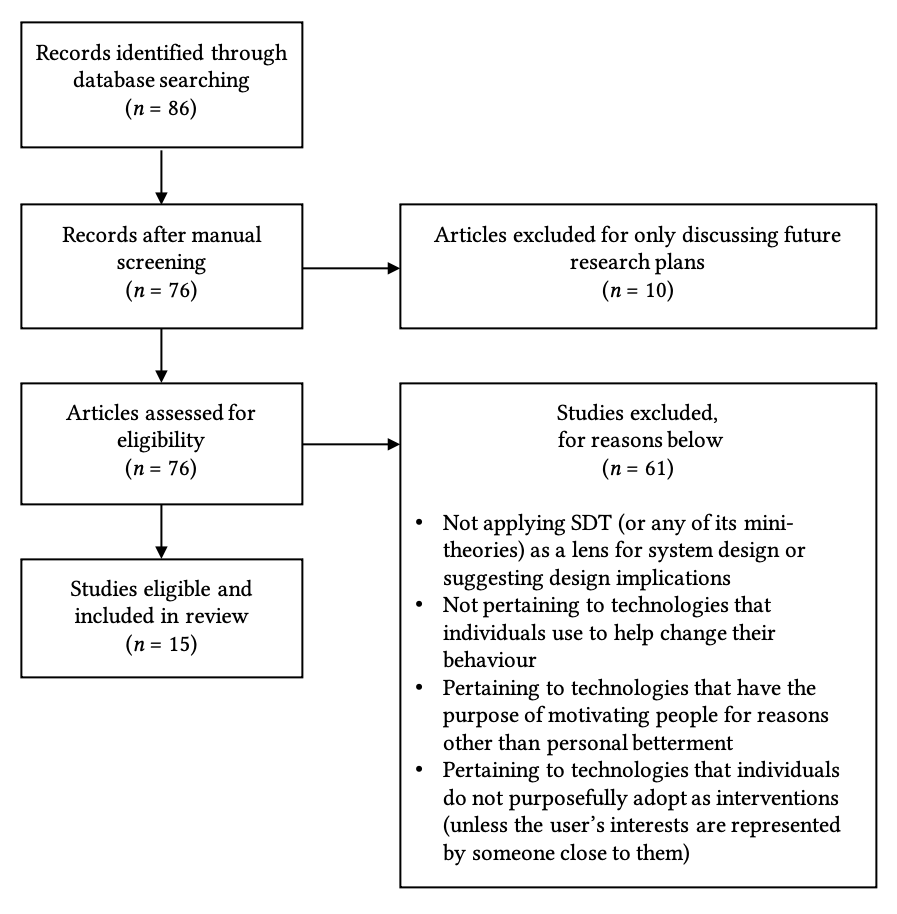}\\
    \caption{Flowchart of the inclusion/exclusion procedure}
    \label{fig:inc}
\end{figure}

After this step, 15 papers remained---see Fig. \ref{fig:inc} for a flowchart of the inclusion/exclusion procedure. The full list of excluded papers and their reasons, as well as our analyses of the included papers, are available on the Open Science Framework.\footnote{\url{https://osf.io/925rn/?view_only=5b8567e524e6416e83ea96fa896948f7}}

\subsection{Coding procedure}
Inspired by Tyack and Meckler \cite{games}, papers were coded with respect to venue, domain, study type (e.g. qualitative, quantitative, mixed), sample size, study duration, methods used, SDT-related measures used, and purposes for citing SDT. For our purposes, we also coded papers based on the user interface (UI) type (e.g., mobile/tablet, wearable, human-robot interaction); the SDT-related constructs that were employed; how those constructs were described; which (if any) of SDT's mini-theories were applied; how the BCT aimed to encourage sustained motivation. We also coded the specific claims made about how any BPNs (autonomy, competence, relatedness) were supported; other claims about how the BCT facilitates an internal PLOC/autonomous regulation; contextual influencing factors that were considered; and the paper's stated novelty or contribution. 

During this process, the lead author read through each paper in its entirety, using different colours to highlight extracts that fit each category. For design recommendations, the coder appealed to the given authors' own claims about the SDT construct they believe it supports, rather than making inferences based on their understanding of the theory. This involved a combination of placing direct extracts from the papers under the relevant headings and making short descriptive summaries. During the next stage, the same author generated codes to capture patterns in design suggestions for supporting specific SDT constructs, and wrote some reflections on the strengths and limitations of each study. The coding spreadsheets are included as supplementary material. 

After this initial phase of analysis, we generated higher-level themes to capture similarities in the design suggestions made by different papers (within the respective BPNs they are meant to support), as well as contextual influencing factors. Whereas most of the generated codes were purely categorical, placing quotes or summaries of content under its corresponding category, based on the authors' own claims, a part of the analysis process was reflexive \cite{braun2019reflecting,clarke2015thematic}, as the coder interpreted overarching themes, and compared the authors' description of the role of each construct with their understanding of Ryan and Deci's \cite{ryan2000intrinsic,ryan2012multiple,ryan2000self,ryan2016facilitating,ryan2017self,ryan2022we} evolving description of the theory and its elements. 

Our reviewing procedure followed that of previous systematic reviews in HCI on interface design and user experience (e.g., \cite{sys1,sys2,sys3,games}), and was guided by Arksey and O'Malley's \cite{arksey2005scoping} methodological framework.

\section{Results}
The following section reports our analysis of the 15 publications we found in HCI venues that apply SDT towards the design of various forms of BCTs (Table \ref{table:papers-included}). Of these, only eleven were full research papers, while three were extended abstracts, and one was a late-breaking report. We decided to include all of these to give a general overview of exploratory work in this area. 

\begin{landscape}
\begin{table}
\footnotesize
\caption{Papers included in review}
\centering
\begin{tabular}[t]{p{0.2cm} p{1.8cm} p{0.5cm} p{6.7cm} p{1.3cm} p{2.3cm} p{1.5cm} p{1.6cm} p{0.6cm}}
\toprule
\# & Authors & Year & Title & ACM Venue & Application domain & UI type & Publication type & Study type\\
\midrule
1 & Aufheimer et al. & 2023 & An Examination of Motivation in Physical Therapy Through the Lens of Self-Determination Theory: Implications for Game Design & CHI & Physical rehabilitation, Health & Unspecified & Full research paper & qual\\
2 & Bomfim \& Wallace & 2018 & Pirate Bri’s Grocery Adventure: Teaching Food Literacy through Shopping & CHI & Education, Health & Mobile & Extended abstract & n/a\\
3 & Chaudhry et al. & 2022 & Formative Evaluation of a Tablet Application to Support Goal-Oriented Care in Community-Dwelling Older Adults & MHCI & Health, Care & Tablet & Full research paper & qual\\
4 & Ferron \& Massa & 2013 & Transtheoretical Model for Designing Technologies Supporting an Active Lifestyle & CHItaly & Health, Fitness & Mobile & Full research paper & quant\\
5 & Ford et al. & 2012 & Self-Determination Theory as Applied to the Design of a Software Learning System Using Whole-Body Controls & OZCHI & Education & Organic UI & Full research paper & qual\\
\addlinespace
6 & Jansen et al. & 2017 & Personas and Behavioral Theories- A Case Study Using Self-Determination Theory to Construct Overweight Personas & CHI & Health, Fitness & Unspecified & Full research paper & qual\\
7 & Lehtonen et al. & 2019 & Movement Empowerment in a Multiplayer Mixed-Reality Trampoline Game & CHI Play & Health, Fitness & Organic UI & Full research paper & mixed\\
8 & Lerch et al. & 2018 & Understanding Fitness App Usage Over Time: Moving Beyond The Need For Competence & CHI & Health,Fitness & Mobile & Extended abstract & quant\\
9 & Molina et al. & 2023 & Motivation to Use Fitness Application for Improving Physical Activity Among Hispanic Users: The Pivotal Role of Interactivity and Relatedness & CHI & Health, Fitness & Mobile & Full research paper & mixed\\
10 & Putnam et al. & 2017 & Effects of Commercial Exergames on Motivation in Brian Injury Therapy & CHI Play & Health, Physical rehab & Organic UI & Extended abstract & quant\\
11 & Saksono et al. & 2020 & Storywell: Designing for Family Fitness App Motivation by Using Social Rewards and Reflection & CHI & Health, Fitness & Mobile, Wearable & Full research paper & qual\\
12 & Sinai \& Rosenberg-Kima & 2022 & Perceptions of Social Robots as Motivating Learning Companions for Online Learning & HRI & Education & HRI, PC & Late-breaking report & quant\\
\addlinespace
13 & Van Minkelen et al. & 2020 & Using Self-Determination Theory in Social Robots to Increase Motivation in L2 Word Learning & HRI & Education & HRI & Full research paper & quant\\
14 & Villalobos-Zúñiga et al. & 2021 & Informed Choices, Progress Monitoring and Comparison with Peers: Features to Support the Autonomy, Competence and Relatedness Needs, as Suggested by the Self-Determination Theory & MobileHCI & Health, Fitness & Wearable, Mobile & Full research paper & qual\\
15 & Yang et al. & 2022 & Magic Brush: An AI-based Service for Dementia Prevention focused on Intrinsic Motivation & CSCW & Health, Care & Tablet & Full research paper & mixed\\
\bottomrule
\end{tabular}
\label{table:papers-included}
\end{table}
\end{landscape}

In terms of user evaluation, the results of 4/15 papers were theoretical: drawing design implications from an interview with experts \cite{Aufheimer}, an interview with envisioned users \cite{Jansen}, watching a video of someone using the BCT \cite{Sinai}, or not involving any participants \cite{Bomfim}. Another two papers drew implications from surveying existing app users once \cite{Molina,lerch}. Of the 9/15 papers that tested a novel BCT design with users, four engaged with the BCT only once \cite{Lehtonen,Putnam,Minkelen,Yangg}. The remaining five studies were done over the course of 6 sessions (of 90 min each) \cite{Fordd}, one month \cite{Villalobos}, 3 months \cite{Saksono}, 6 months \cite{Chaudhry}, and two years \cite{Ferron} respectively. The maximum sample size was 211, with a median of 35 across studies. 

\subsection{Application domains}
We identified two primary domains of application for BCTs, with some overlap. The most prominent was \textit{health}, which pertained to 12/15 of the papers, and the other was \textit{education} (4/15). Health was coupled with the subcategories of \textit{fitness} (7/15), \textit{physical rehabilitation} (2/15), and \textit{care} (2/15), and once with \textit{education} (i.e., food/nutrition literacy). 

\subsection{UI types}
In total, the 15 papers developed and/or evaluated six different types of interfaces, again with some overlap (e.g., software that can either be implemented on different devices, or BCTs that involve multiple devices at once). The types were \textit{mobile, wearable, tablet, PC, HRI, and Organic UI.} While some applications may have been compatible with multiple devices (e.g. mobile phones, tablets, laptops, PCs), we did not make such inferences ourselves. We used \textit{mobile} to refer to applications that were described as primarily intended for mobile phones, and \textit{tablet} when the application utilised tablet features (e.g., drawing on the screen with a stylus). We used \textit{PC} when the tool specifically required a large screen and/or keyboard typing (e.g., some educational programs). We used \textit{HRI} (human-robot interaction) to refer to BCTs that involved social interaction with a robot, and \textit{organic UI} for interfaces that involved users' whole bodies (e.g., as in certain exercise games). 

\textit{Mobile} was the most prominent UI type (4/15), followed by \textit{organic UI} (3/15), whereas \textit{wearable}, \textit{tablet}, \textit{PC}, and \textit{HRI} were each explored in 2/15 papers respectively. Two papers were implementation-neutral/did not specify a specific UI type. 

\subsection{Which of SDT's theories and constructs were applied (RQ1)}

Of the reviewed papers, only 3/15 explicitly mentioned the mini-theories they applied, although all employed constructs from specific mini-theories. 13/15 employed the BPNs (as put forward in \textit{Basic Psychological Needs Theory}, BPNT), of which one only used the relatedness construct \cite{Saksono}.  11/15 employed intrinsic vs extrinsic motivation (as put forward in CET), and 5/15 employed the different motivation types of the autonomy-control continuum (combining CET with OIT) \cite{Aufheimer,Ferron,Jansen,Villalobos,Molina}. One paper employed constructs drawn from the Intrinsic Motivation Inventory \cite{imi} instead of the BPNs \cite{Putnam} , and one employed self-determination as a construct instead of intrinsic motivation \cite{Chaudhry}.

Of those that mentioned the application of specific mini-theories, all three said they used BPNT \cite{Aufheimer,lerch,Minkelen}, although Aufheimer \textit{et al.} \cite{Aufheimer} combined it with OIT, and Van Minkelen \textit{et al.} \cite{Minkelen} combined it with CET. 

\subsubsection{How was the role of the constructs in facilitating autonomous regulation described?}

Most papers offered a short description of the BPNs (14/15), and one employed them without summarising what each need means/entails \cite{Bomfim}.\footnote{The format of the paper was an extended abstract, which may explain the oversight.}  

Several papers explained that satisfying these BPNs is required for/can facilitate the internalisation of \textit{motivation} \cite{Bomfim,Saksono,lerch}, the \textit{behaviour }\cite[p.2]{lerch,Aufheimer}, or the \textit{performance of an activity} \cite{Molina}.  Several others maintained that satisfying the BPNs will enhance \cite{Putnam,Fordd,Jansen,Yangg}, maintain \cite{Minkelen}, or increase \cite{Sinai,Jansen} \textit{intrinsic motivation}: e.g., ``When these three needs are met, intrinsic motivation is expected to increase'' \cite[p.1045]{Sinai}. One simply stated that satisfying the needs is necessary for ``optimal function and growth'' \cite[p.23]{Lehtonen}, and another that it leads to self-determination \cite{Villalobos}.

As to how the internalisation process works, some understood it as a matter of likelihood: the more of the BPNs are satisfied, the greater the chance that motivation will fall on the more autonomous side of the continuum \cite{Molina, Jansen}, e.g.,: ``If all three of these basic psychological needs are sufficiently met, intrinsic motivation to lead a healthy lifestyle is probably high'' \cite[p.2128]{Jansen}. Others maintained that moving along the continuum happens over time, as a staged/gradual process---either by satisfying the needs \cite{Saksono,Yangg}, or ``by providing personalized feedback to people with different motivational levels or at different stages of the behavior change process'' \cite[p.2]{Ferron}. The latter argued that, thereby, a person may finally ``reach'' intrinsic motivation: ``[SDT is a] theory describing behavioral change as a gradual process starting from motivation ... toward increasing levels of internal regulation, finally reaching intrinsic motivation'' \cite[p.1]{Ferron}. 

In defining the BPNs, descriptions seemed generally consistent with how Ryan and Deci \cite{ryan2000self,ryan2017self} describe them. However, only rarely (5/15) were the particular \textit{roles} of the constructs described in terms of how they might facilitate the integration process from more external to internal sources of motivation, as OIT describes \cite{ryan2000self,ryan2017self}. For \textit{competence} in particular, several summaries may be construed as overly reductive through that lens: e.g., defining competence simply as ``the self-perceived ability to learn new things and receive feedback'' \cite[p.370]{Minkelen}; as having mastery over challenging tasks \cite[p.23]{Lehtonen}
; as a feeling of confidence in one's ability \cite{Molina,Saksono,Putnam,Jansen,Yangg}; or as a feeling of effectance from receiving positive feedback \cite{Sinai,Yangg}. In contrast, only five papers \cite{Aufheimer,Molina,Villalobos,Bomfim,Putnam} engaged with role of competence support to facilitate ``higher-order reflection''  on the person's values, goals, and purposes (an essential part of the integration process \cite{ryan2017self}), more than simply feeling sufficiently challenged or confident due to receiving positive feedback. This could be because most papers drew the constructs from BPNT or SDT in general, rather than, at least explicitly, considering the particular role the constructs play in the mini-theories in facilitating the internalisation of regulation. It could also be because the remaining (10/15) papers only employed constructs from CET (i.e., extrinsic vs intrinsic motivation), rather than OIT's different motivation types, and the process of moving between them. 


\subsection{Reasons for applying SDT (RQ2)}


All of the reviewed papers suggested sustained motivation as a potential benefit of applying SDT. Some described it in terms of enhancing motivation for a given behaviour \cite{Lehtonen, Bomfim, Ferron, Jansen}, increasing engagement with the BCT \cite{Aufheimer,lerch,Saksono,Fordd,Yangg}, improving the quality of their experience with the BCT \cite{Sinai}, or supporting users' sense of self-determination or empowerment \cite{Aufheimer,Bomfim,Chaudhry,Villalobos}. Holistic benefits like promoting users' overall well-being  \cite{Aufheimer,Chaudhry,lerch} or supporting a healthy lifestyle \cite{Jansen,Ferron,Saksono} were also mentioned. 

Most papers referred to ``intrinsic motivation'' as a part of the defining aim of employing SDT (10/15), as opposed to the integrated regulation of a specific goal or value of a behaviour (3/15) \cite{Saksono,Aufheimer,Molina}. This owes to the fact that, as mentioned, rather than describing motivation as a movement between different \textit{types} or orientations to motivation, constructs from CET and OIT were often conflated. As such, intrinsic motivation was sometimes understood as something that is already present and merely increases \cite{Jansen,Fordd,Sinai,Putnam,Yangg}: ``SDT proposes that intrinsic motivation of the users of a system such as Stomp is enhanced by meeting three key needs'' \cite[p.147]{Fordd}, or ``When these three needs are met, intrinsic motivation is expected to increase'' \cite[p.1045]{Sinai}. One paper acknowledged the different types of motivation on the autonomy-control continuum, but argued that all of those in the extrinsic category in OIT will fail to sustain long-term behaviour change: ``SDT introduces a control–autonomy continuum ... from amotivation (or absence of intention to act) to external regulation (to obtain a reward) to introjected regulation (to avoid guilt) to identification (accepted external regulation) to integration (self-determined action). Additionally, the SDT explains that these ... extrinsic motivation types can urge a person to behave a certain way in the short-term but will fail to maintain the behavior over more extended periods.'' \cite[p.2]{Villalobos}.

Application aims were typically phrased in terms of sustained usage of the intervention (e.g., increasing/enhancing  the ``motivation to use a health app'' \cite[p.1]{Saksono} or ``strengthening user engagement'' \cite[p.149]{Fordd}), rather than promoting the activity that the BCT promotes outside of the intervention (making \textit{exercise} more inherently motivating, e.g., by helping users internalise its value to their lives). Even when the aims were explicity stated as motivating users to engage in healthy behaviours, it was implied that this would be done \textit{via} motivating them to use the BCT, rather than helping them internalise the value of the activity \textit{per se}. For example:  ``the satisfaction of the basic needs through the use of technology of technology should result in intrinsic motivation to use the fitness App'' \cite[p.4]{Molina}. Only one paper made explicit mention of helping users sustain positive behaviours in their lives beyond the BCT \cite{Bomfim}. That is, to ``develop an understanding of the nutritional benefits of foods ... [in order to] learn and internalize content to maintain their outside the game'' \cite[p.1-6]{Bomfim}.  In this vein, one paper made it explicit that enhancing engagement with the app can have the effect over time of helping them engage in the target behaviour, even if it will not necessarily enhance their intrinsic motivation for the behaviour itself \cite{Molina}: ``Importantly, behavior is not always driven by intrinsic motivation (e.g., using fitness Apps due to sheer pleasure). In fact, behavior related to PA [physical activity] is mostly driven by extrinsic motivational factors ... Integrated regulation is the nearest to intrinsic motivation and is desired for the sustained use of fitness Apps. Sustained use of the device could result in long-term PA adherence.'' \cite[p.3-4]{Molina}. Two papers considered the BCT as a supplementary tool for in-person therapy, and argued that, thus, applying SDT to the BCT will only partly help sustain their motivation \cite{Aufheimer}.

\subsubsection{How the BCT aims to encourage sustained motivation}

We also analysed papers in terms of their claims regarding how the BCT might encourage sustained motivation. This was often stated as the outcome of satisfying one or more BPN in some way (13/15), as it \textit{enhances} \cite{Saksono}, \textit{stimulates} \cite{Minkelen}, \textit{strengthens} \cite{Yangg} or \textit{increases} \cite{Sinai} users' intrinsic motivation; supports their sense of self-determination \cite{Villalobos,Chaudhry}; or makes the BCT more enjoyable \cite{Ferron,Lehtonen,Yangg} (e..g, `trying to make users experience sensory pleasure' \cite[p.4]{Ferron}. Chaudhry \textit{et al.} \cite{Chaudhry} also suggested avoiding common reasons for abandonment (e.g., a lack of trust and familiarity with the BCT), and supporting re-engagement. In the context of exergames, 3/15 papers also suggested that building user confidence, in particular, will motivate sustained use \cite{Putnam,lerch,Lehtonen}. In the context of therapy, 2/15 papers maintained that making BCTs more engaging will not sustain motivation as well as in-person therapy on its own, but can facilitate the process \cite{Aufheimer,Putnam}.



\subsection{How SDT was applied to inform design decisions (RQ3)}

Beyond the description of SDT and its concepts, the meanings of the relevant constructs were further evidenced in how they were translated into design requirements and operationalised in measures. In analysing the reviewed work, we distinguished between \textit{claims about supporting each of the BPNs by design}, and\textit{ claims about facilitating an internal PLOC/autonomous regulation}, as both are important for internalising motivation \cite{metantoumanis,ryan2017self,ryan2000self}. Finally, we discuss the use of any SDT-related measures in validating the different design suggestions. 

\begin{figure}
    \centering
    \includegraphics[scale=0.8]{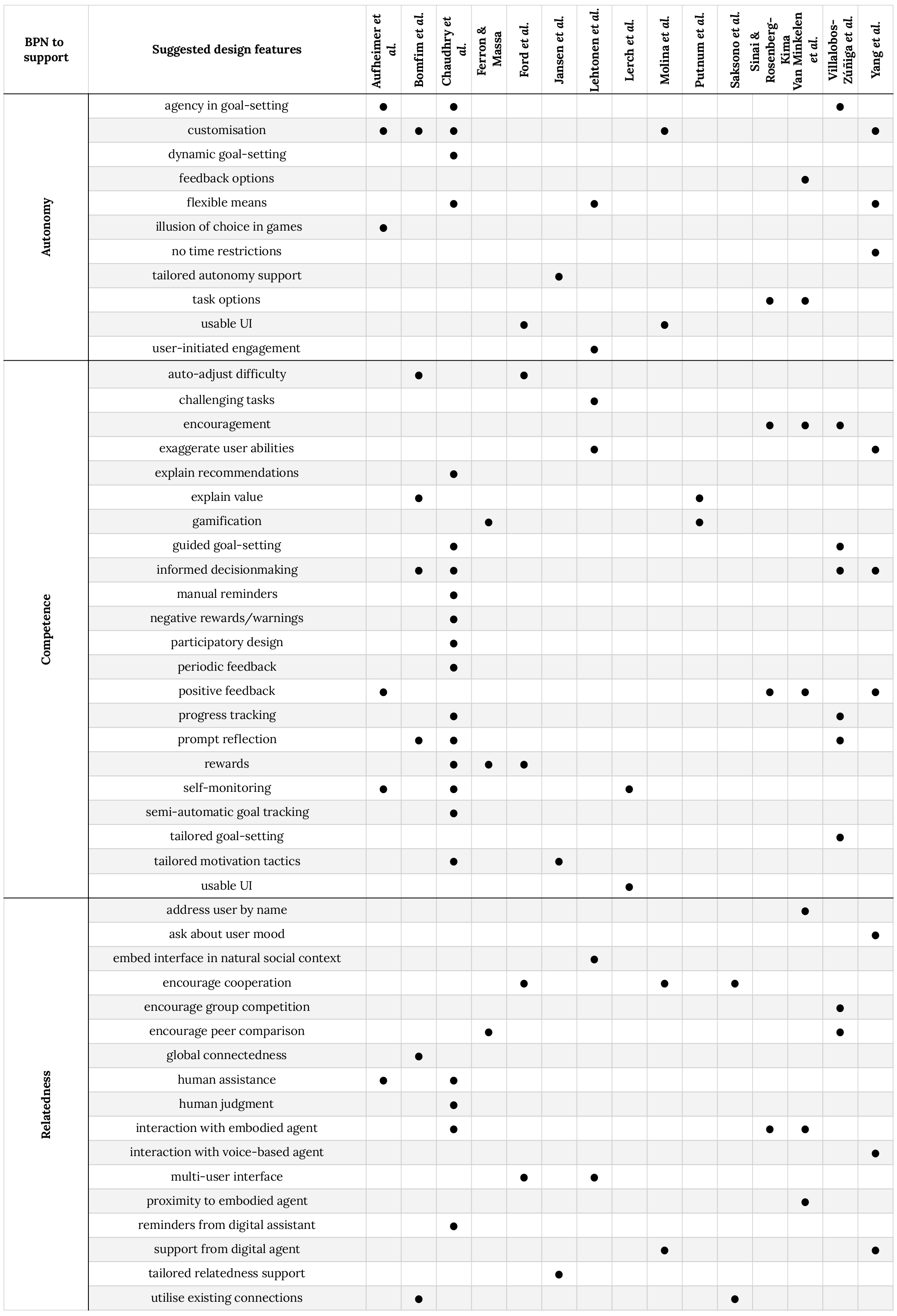}\\
    \caption{Design suggestions for supporting users' three \textit{basic psychological needs} \cite{ryan2000self} in behaviour change technologies}
    \label{fig:codes}
\end{figure}

Regarding supporting the BPNs, \textit{competence} received the most treatment (13/15), followed by \textit{relatedness} (12/15) and \textit{autonomy} (10/15)---see Fig.\ref{fig:codes}.

\subsubsection{Supporting autonomy by design}
We captured the range of design suggestions for supporting user `autonomy' with the themes \textit{supply choice in task/outcome}, \textit{supply choice in means}, e\textit{nhance UX/usability}, \textit{afford ability to voice}, and \textit{tailor features to individual differences}. The 11 codes underlying these themes, and their prevalence across papers, are shown in Fig.\ref{fig:codes}.

\textbf{Supply choice in task/outcome}: Multiple papers suggested that the need for autonomy could be supported by increasing the user’s degree of choice over the tasks they completed, and/or the outcomes they achieved. We captured these under the codes \textit{agency in goal setting} in 3/15 papers, \textit{task options} (2/15), and \textit{dynamic goal-setting} (1/15).  \textit{Agency in goal-setting} ranged from allowing users to set their own goals (e.g., their weekly step-goal \cite{Villalobos}), to selecting their own goals from an example goal list \cite{Chaudhry}. Giving users full autonomy over their goals was considered more appropriate in some contexts than others. For instance, in expert interviews, Aufheimer \textit{et al.} found that physical therapy is often ``a space devoid of meaningful provision of autonomy'' \cite[p.13]{Aufheimer} as patients do not possess the relevant expert knowledge, but maintained that they should still be involved in ``meaningful and collaborative goal-setting'' with therapists. \textit{Task options} referred to features that offer the user some choice over how they want to execute the task, either by giving options to choose from (e.g., do you want to do X or Y? \cite{Sinai} or allowing them to manipulate some small aspect of the format of the task without changing the nature of the task 
 \cite{Minkelen}. Finally, \textit{dynamic goal-setting} referred to allowing users to adjust and update their goals over time \cite{Chaudhry}.

\textbf{Supply choice in means} captured suggestions for offering some degree of freedom in \textit{how} tasks/outcomes were done/achieved. This included the codes \textit{customisation} (5/15), \textit{flexible means} (3/15), \textit{user-initiated engagement} (1/15), and \textit{no time restrictions} (1/15). \textit{Customisation }included allowing users to help shape the content of the program \cite{Yangg} and offering means for users to specify their unique needs and preferences---either from pregiven options \cite{Bomfim,Molina}, or through more qualitative means like motivational interviewing \cite{Chaudhry} or manually calibrating the BCT to individual traits and abilities. \textit{Flexible means} referred to giving the user a task or action plan that they are free to implement according to their needs, circumstances and abilities \cite{Chaudhry, Lehtonen,Yangg}. \textit{Illusion of choice in games} referred to a suggestion for gamifying a physical therapy task plan such that users have an ``illusion of autonomy'' even if outcomes were set \cite{Aufheimer}. Finally, \textit{no time restrictions} meant allowing users to take the time they want to complete a task \cite{Yangg}.

\textbf{Enhance UX/usability}: Two papers argued that autonomy could be supported by making the interface more intuitive and easy to use. The code \textit{usable UI} (2/15) included the suggestion that an interface that's easy to navigate will allow users to freely exercise their options \cite{Molina}, as well as the suggestion that a whole-body interface that allows users to behave in intuitive (organic) ways will likewise support their autonomy \cite{Fordd}.  

\textbf{Afford ability to voice}: This theme captured a suggestion for giving users a degree of autonomy by allowing them to voice their experience with the BCT. The code \textit{feedback options} (1/15) referred to a suggestion for offering options for users to give feedback on how they found a task \cite{Minkelen}. In particular, ``the need for autonomy was also supported by presenting a green smiley which the child could press to start the lesson, and a red smiley which the child could press to indicate he/she had not understood the instruction'' \cite[p.371]{Minkelen}.

\textbf{Tailor features to individual differences}: Finally, this theme captured a suggestion for incorporating considerations of SDT support in user-centred design methodology. The code \textit{tailored autonomy support} (1/15) referred to a suggestion that designers should anticipate variations users' needs for autonomy at the design stage (captured as nuances in persona descriptions), so as to include appropriate features, whatever they may be, tailored to different users' relative needs for autonomy support \cite{Jansen}.

\subsubsection{Supporting competence by design}
We captured 20 codes relating to suggestions for supporting user `competence'. We discuss these under the themes \textit{adjusting difficulty}, \textit{giving validation}, \textit{offering incentives}, \textit{improving self-knowledge},\textit{ improving task knowledge}, \textit{increasing user involvement}, \textit{enhancing UX/usability}, and \textit{tailor features to individual differences}. 

\textbf{Adjusting difficulty}: This theme captured suggestions that related to making tasks either more difficult or easier, or merely appear as such, so that users feel competent executing them. It included the codes \textit{auto-adjust difficulty} (2/15), \textit{exaggerate user abilities} (2/15), \textit{tailored goal-setting} (1/15), and  \textit{challenging tasks} (1/15). \textit{Auto-adjust difficulty} was an implementation of the idea in SDT that challenges should neither be too difficult nor too easy, and so it should adjust according to what users are able to achieve \cite{Bomfim,Fordd}. \textit{Exaggerate user abilities} refers to making users feel more powerful by apparently increasing their abilities, such as by showing them a simulated version of themselves with exaggerated performance \cite{Lehtonen}, or digitally enhancing things they produce \cite{Yangg}. \textit{Tailored goal-setting} meant adjusting goals based on the user's recent performance \cite{Villalobos}, and \textit{challenging tasks} just meant to make tasks extra challenging to be more motivating \cite{Lehtonen}. 

\textbf{Giving validation}: Several papers suggested making users feel validated in their efforts as a way to support their sense of competence. This theme captured the codes \textit{positive feedback} (4/15), \textit{encouragement} (3/15), \textit{periodic feedback} (1/15). \textit{Positive feedback} included references to giving constructive feedback to users, either depending on whether or not they succeeded \cite{Sinai, Minkelen}, or regardless: ``the virtual assistant gives positive feedback to the image that the user draws regardless of the user’s performance. The positive feedback can increase the competence ... of users even if they are not good at drawing'' \cite[p.6]{Yangg}. \textit{Encouragement} was suggested to help users not feel incompetent when they struggle \cite{Sinai, Minkelen, Villalobos}). For instance: ``It seems like the answer is wrong, but I know you can do it. You just need some more time to think about it'' \cite[p.1046]{Sinai}. Finally, \textit{periodic feedback} was the suggestion of offering users a weekly report giving feedback on their achievements/progress that week \cite{Chaudhry}.

\textbf{Offering incentives} captured suggestions for supporting users' needs for competence by offering them some incentive to keep on track with their goals. This included \textit{rewards} (3/15), \textit{gamification} (2/15), and \textit{negative rewards/warnings} (1/15). \textit{Rewards} referred to offering some form of virtual rewards\cite{ Chaudhry,Ferron,Fordd}) whilst \textit{negative rewards/warnings} referred to virtual indicators of a lack of progress \cite{Chaudhry}. \textit{Gamification} techniques were suggested to make tasks more engaging (for being both challenging and rewarding) \cite{Putnam,Ferron}.

\textbf{Improving self-knowledge}: Some papers suggested that competence can be supported by helping users notice things regarding their progress and choices. This captured the codes \textit{prompt reflection} (3/15), \textit{ self-monitoring} (3/15), and \textit{progress tracking} (2/15). \textit{Prompt reflection} refers to showing information or adding some level of friction to prompt users to reflect on things like the impact of their past goals \cite{Chaudhry}, how contextual factors may have affected their progress \cite{Villalobos}, or giving them a chance to change their decisions \cite{Bomfim}. \textit{Self-monitoring} refers to showing information to help users monitor their past achievements/goals they have already reached \cite{Aufheimer, Chaudhry,lerch}, while \textit{progress tracking} refers to showing information to help users see how far they are from reaching their current goal(s) \cite{Chaudhry,Villalobos}.

\textbf{Improving task knowledge}: There were a few suggestions for supporting competence by helping users better understand what they are asked to do. This included the codes \textit{informed decisionmaking} (4/15),\textit{ guided goal-setting} (2/15), \textit{explain value} (2/15), \textit{explain recommendations} (1/15), and \textit{semi-automated goal tracking} (1/15). \textit{Informed decisionmaking} had to do with offering users more information with options, such as what they can expect from choosing a given option \cite{Yangg, Villalobos}, what its particular benefits are \cite{Bomfim} or why a given option may be relevant to the user in particular \cite{Chaudhry}. \textit{Guided goal-setting} meant offering users help with setting their personal goals, either through an example goal list  \cite{Chaudhry} or suggest adaptable goals to each individual’s ability level \cite[p.9]{Villalobos}. \textit{Explain value} refers to showing information on \textit{why} certain tasks or goals are meaningful or useful, so as to help users internalise their value \cite{Bomfim, Putnam}. \textit{Explain recommendations} meant explaining why recommendations are relevant to the particular user \cite{Chaudhry}. Finally, \textit{semi-automated goal tracking} was suggested as a way to minimize burden on the user whilst still affording them a level of autonomy \cite{Chaudhry}.

\textbf{Increasing user involvement}: One paper suggested that competence can be supported by letting a user play an active role in the BCT's mechanism for motivating them. The code \textit{manual reminders} (1/15) captured the idea that users would feel more competent if they set their task reminders themselves (in an app) \cite{Chaudhry}.

\textbf{Enhancing UX/usability}: One paper suggested that users would feel more competent if the interface is easy to use. The code \textit{usable interface}captured the idea of improving the usability of the platform overall, such that users have a greater sense of efficacy when using it \cite{lerch}.\footnote{Whereas Fordd \textit{et al.} \cite{Fordd} suggested that usability supports user autonomy, Lerch \textit{et al.} \cite{lerch} suggested it supports user competence. We report on how the authors employed the tactics.}

\textbf{Tailor features to individual differences}: Finally, three papers suggested individual differences may determine how competence is best supported. This captured the codes \textit{tailored motivation tactics} (2/15), and \textit{participatory design} (1/15). Papers that advocated \textit{tailored motivation tactics} suggested that the most useful/effective tactics for supporting user competence depend on contextual factors like the stage of adoption users are at (i.e., whether they just starting on a goal or already making progress \cite{Ferron}, or their personality \cite{Jansen}. \textit{Participatory design} meant involving end-users in the interface's design (particularly visualising their progress) such that it is easy for them to understand \cite{Chaudhry}.

\subsubsection{Supporting relatedness by design} 
We generated 17 codes relating to suggestions for supporting user `relatedness', captured under the themes \textit{interaction with conversational agent}, \textit{user connectedness}, \textit{interaction with external people}, \textit{interaction with human expert}, \textit{considerate gesture from interface}, and \textit{tailor featues to individual differences}.

Interestingly, the most prominent theme here was \textbf{relate to conversational agent}, rather than connecting users with other people. This included the codes \textit{interaction with embodied agent} (3/15), \textit{reminders from digital assistant} (1/15), \textit{proximity to embodied agent} (1/15), \textit{support from digital agent} (2/15), and \textit{interaction with voice-based agent} (1/15). The code \textit{interaction with embodied agent} referred to suggestions to let users engage in human-like social interactions with a social robot, whether for care, encouragement and support \cite{Sinai, Minkelen} or merely for assistance \cite{Chaudhry}. It was also suggested that receiving \textit{reminders from digital assistant} (i.e., a virtual AI agent \cite{Chaudhry}) would make users feel more related,  as would receiving \textit{guidance from digital agent} (e.g., feedback such as``You did a great job'' and ``I love the picture.'' \cite{Yangg}, or personalised suggestions and help \cite{Molina}). Under \textit{interaction with voice-based agent}, Yang \textit{et al.} used a celebrity voice as a digital assistant of a virtual therapist: ``Several studies have shown that not only do people often feel connected and involved with celebrities and others appearing in the media but the sense also develops'' \cite[p.5]{Yangg}. Finally, \textit{proximity to embodied agent} referred to the suggestion that a social robot's mere presence is enough support a sense of relatedness \cite{Minkelen}.

\textbf{Relate to other users}: Several papers suggested supporting relatedness by encouraging users to connect with other users in some regard. This included the codes \textit{multi-user interface} (2/15), \textit{encourage cooperation} (3/15), \textit{encourage peer comparison} (2/11), \textit{encourage group competition} (1/15).  \textit{Multi-user interface} referred to exergames that require multiple players so as to make players feel more connected \cite{Fordd,Lehtonen}. \textit{Encourage cooperation} referred to suggestions for encouraging users to collaborate with other users, either by tasks that require cooperation\cite{Fordd,Saksono}, or by visualising data of user achievements in collaborative ways \cite{Molina}.  \textit{Encourage peer comparison }meant features for showing the progress/achievements of similar users \cite{Ferron,Villalobos}. However, in finding this to backfire, \cite{Villalobos} suggested \textit{encourage group competition} as an alternative (i.e., letting users compete in teams to forge a sense of comradery).

\textbf{Relate to external people}: Suggestions here pertained to encouraging users to relate to other (non-user) people in the world. This included the codes \textit{utilise existing connections} (2/15), \textit{global connectedness} (1/15), and \textit{embed interface in natural social context} (1/15). \textit{Utilise existing connections} referred to suggestions for features that encouraging users to reach out to their existing friends or family  \cite{Sinai,Minkelen}. \textit{Global connectedness} applied to a suggestions for informing the user about broader initiatives that partake in similar behaviours, so as to foster a sense of community \cite{Bomfim}. Finally,\textit{ embed interface in natural social context} was the suggestion that playing a mixed-reality exergame in a natural social setting will encourage social interaction, and, hence, support relatedness \cite{Lehtonen}. 

\textbf{Interaction with human expert}: This captured suggestions regarding the importance of human actors (particularly in the context of care/therapy) for supporting people's sense of relatedness. This included the codes \textit{human assistance} (1/15) and\textit{ human judgment} (1/15). The code \textit{human assistance} referred to a suggestion that physical therapy games should only be considered a supplementary technology to human-led therapy, as ``the human relationship between therapist and patient serves as a source of enjoyment and comfort that should be respected and supported by technical interventions'' \cite[p.13]{Aufheimer}. Similarly, Chaudry \textit{et al.} suggested that \textit{human judgment} should be used to help personalise goal suggestions, and that doing so will support user relatedness in some way: ``collected information [about user needs, values and preferences] is sent to the care manager’s portal, who is then able to send goal suggestions to the participants’ app (relatedness)'' \cite[p.6]{Chaudhry}. 

\textbf{Considerate gesture from interface}: Two papers suggested that interfaces that perform supportive humanlike social gestures to foster a sense of relatedness. It contained the codes \textit{address user by name} (1/15) \cite{Minkelen} and  \textit{ask about user mood} (1/15) \cite{Yangg}. \textit{Ask about user mood} referred to the idea of asking a user how they feel that day so as to  maintain ``healthy emotional conditions'' \cite[p.5]{Yangg}.

Finally, \textbf{tailor features to individual differences} captured a suggestion that the appropriate relatedness supporting features may depend on the user. It contained the code \textit{tailored relatedness support} (1/15), which described a suggestion to anticipate variations users' needs for relatedness at the design stage so as to determine which approach(es) are most suitable \cite{Jansen}.

\subsubsection{Facilitating an internal perceived locus of causality/autonomous regulation by design}

Beyond supporting the BPNs, 10/15 papers considered how to support an internal PLOC/autonomous regulation by design. Here we included any claims about facilitating the internalisation process towards greater self-determination, without mention of supporting a specific need (even if they may implicitly relate to one). 

Suggestions here included creating space in which patients can be involved in meaningful goal-setting \cite{Aufheimer}; helping users internalise content \cite{Bomfim}; allowing increasing users' awareness of their progress \cite{Chaudhry,Ferron,Villalobos}; providing tailored feedback at different stages of the process \cite{Ferron}; by identifying specific behavioural variables of different users in the design stages \cite{Jansen}; promoting interaction with other users \cite{Molina}; helping the user feel valued and appreciated \cite{Minkelen}; clearly communicate the BCT's value \cite{Putnam}; and extending the frequency of satisfying \cite{Saksono} or educational \cite{Bomfim} moments over time.



\subsubsection{Measures used}
Of all the papers we reviewed, only 4/15 used full validated SDT-related scales to measure SDT-related constructs (i.e., the \textit{Ubisoft Perceived Experience Questionnaire (UPEQ)} \cite{Lehtonen}; the \textit{Technology-based Experience of Need Satisfaction questionnaire} \cite{Peters2018-ra}; the \textit{Balanced Measure of Psychological Needs (BMPN) scale} \cite{sheldon2012balanced}; and the \textit{Perceived Intrinsic Motivation (PIM) Questionnaire} \cite{Sinai}. Another 2/15 papers used shortened versions of existing scales: a reduced (Italian) version of the \textit{Sport Motivation Scale} \cite{Ferron}, and a shortened version of the \textit{Intrinsic Motivation Inventory (IMI) }scale \cite{Putnam}.  \cite{Yangg} measured intrinsic motivation with three statements developed by \cite{davis1992extrinsic}.

Some papers (3/15) did not use SDT-related scales or questionnaires, but still based their data analysis on the theory. This included using SDT to inform interview topics/code participant responses \cite{Jansen}, asked researchers familiar with SDT to evaluate app features through an SDT lens \cite{Villalobos}, or made their own scale for measuring participants' level of engagement with a task \cite{Minkelen}.

4/15 papers used no SDT-related measures \cite{Bomfim, Chaudhry, Fordd, Saksono}. Instead, these papers based the design of a BCT or prototype on SDT, but either did not test it with users \cite{Bomfim} or measured constructs of user experience/satisfaction unrelated to SDT \cite{Chaudhry,Saksono}. One paper observed children engaging with an exercise game for 90 minutes and made inferences about its ability to support users' BPNs without asking users directly \cite{Fordd}.

\subsection{Contextual determining factors (RQ4)}

Given the breadth of domains and behaviours for which BCTs can be designed, there is some variation in \textit{how} and \textit{to which extent} it makes sense to support each of the three BPNs. Moreover, other theories like the transtheoretical model of behaviour change \cite{trans} suggest that different groups of people will vary in their needs and the interventions that effectively support them. In our review, we found some acknowledgement of factors that may affect the contextual suitability of certain design features and motivation mechanisms.

One contextual factor that was raised was the relative \textit{stage} of behaviour change that an individual was in, e.g., whether they are only just adopting a new behaviour or trying to sustain it \cite{Ferron}. According to Ferron and Massa \cite{Ferron}, new adopters may require more extrinsic motivation (e.g., gamification techniques) and more education, whereas users at the later stage may benefit from ``other incentives that leverage their intrinsic motivation to a healthy and active lifestyle'' \cite[p.5]{Ferron}. Another paper considered individual differences, arguing that designers should ``create complex, yet engaging and highly realistic personas that make [variations in] users’ basic psychological needs explicit'' \cite[p.2127]{Jansen}. Related to this was age: some papers dealt with determining the  appropriate features for users that have diminished autonomy like children \cite{Saksono,Fordd, Minkelen}, or other special requirements like older adults \cite{Chaudhry}. Chaudry \textit{et al.}\cite{Chaudhry} found that their participants (adults >55) found the use of `negative rewards'/warnings motivating, contrary to findings with younger participants \cite{consolvo2008activity}. In the context of exergames for brain injury therapy, \cite{Putnam} found that males had significantly higher intrinsic motivation scores than females, which they took to suggest that therapists `may need to scaffold their female patients more in exergame therapies'' \cite[p.56]{Putnam}. However, given the relatively small sample sizes of studies (n=between 5 and 211), and the particular conditions of each study, the generalisability of these findings are questionable. 

Regarding cultural differences, Molina \textit{et al.} \cite{Molina} investigated the experience of Hispanic users of fitness apps. They maintain that most studies testing the efficacy of mHealth technologies have been conducted with a predominantly white population, and fail to adequately address the needs of non-white communities: ``Hispanics can view PA as `a waste of time,' hold different norms regarding weight and body shape, and value social support with close family ties and obligations. Such culture-specific values may detract Hispanics from engaging with fitness Apps for improving PA' \cite[p.1]{Molina}. To address this, they argue that customisation options should not just account for personal information like age or gender,  but also to the user’s norms, values, and worldviews, which can diverge even within a given cultural group. They also emphasise that the particular implementation and embodiment of features matter.

Another consideration was domain: Aufheimer \textit{et al.} \cite{Aufheimer} maintained that, in applications that require some level of expert regulation (e.g., healthcare), it may not be in users' best interests to give them as much control as in other domains (e.g., fitness), but that a sense of autonomy could still be supported through \textit{collaborative} goal-setting, and games with some illusion of freedom towards predefined outcomes. Lerch \textit{et al.} \cite{lerch} also highlight the importance of external contextual factors, as factors like no longer finding exercise necessary or getting a gym membership can also contribute to app abandonment, and so it may not necessarily mean the BCT was not effective in its purpose. 

\section{Discussion}
Self-determination theory (SDT), an evidence-based theory of human motivation and growth, is one of the major motivational theories in mainstream psychology \cite{metagillison,metantoumanis,ryan2022we}, and has recently gained popularity in HCI \cite{Shaping,Peters2018-ra}. It is particularly useful for 
understanding factors that either support or undermine an individual's ability to sustain their self-determination to engage in certain behaviours, which we believe may usefully guide BCT designers in facilitating sustained behaviour change.
To explore this opportunity, we systematically reviewed 15 HCI papers that employ SDT specifically towards the design of BCTs to explore (\textbf{RQ1}) which of SDT's theories have been applied in HCI towards BCT design, (\textbf{RQ2}) what reasons were cited for this application, (\textbf{RQ3}) how SDT-related constructs have been translated into specific design features/suggestions, and (\textbf{RQ4})  what contextual factors affect the suitability of these design features for different BCTs. 

Whilst SDT has gained popularity in multiple areas in HCI---especially games research---we found that its application towards enhancing user motivation in the context of behaviour change (at least in the HCI community) is still relatively sparse. Our review of papers in the ACM Digital Library containing `self-determination theory' in the abstract returned 84 results. Of these, only 15 were research papers that leveraged the theory to make design suggestions for technologies towards helping individuals change their behaviour in some personally-desired way. All of these papers used key constructs from SDT (such as intrinsic motivation and the basic psychological needs for autonomy, competence, and relatedness) as a basis for proposing design guidelines for BCTs, in a range of application domains. 

To evaluate the progress and remaining opportunities regarding research in this area, we were interested in understanding how exactly SDT was leveraged: the aims for which the theory was applied, how the roles of the theory's constructs were understood, and how this guided the design approach and evaluation of the BCTs. We identified two overarching domains of application: \textit{health} (subsuming \textit{fitness}, \textit{physical rehabilitation} and \textit{care}), and \textit{education}. This involved a range of different interface types: from mobile apps, to human-robot interaction, to wearables (or some combination). 

All papers made design suggestions for supporting at least one of the basic psychological needs that SDT proposes, with most of the focus on `competence'. In our analysis, we generated themes to capture patterns in design solutions that have been proposed across BCT domains and interface types. We also identified other claims about supporting an internal PLOC/autonomous regulation by design, and some contextual factors that may affect the relative suitability of design solutions.

\subsection{How SDT can help BCTs facilitate sustained motivation change}

From our analysis, we identified two primary ways in which SDT has been applied to the design of BCTs, revealing subtly different conceptions about the (a) ultimate aim of BCTs, and (b) how they may help users achieve sustained behaviour change. Approaches may also implicitly contain some combination, although this distinction was not typically acknowledged. 

The first is understanding a BCT as something that makes certain target behaviours more intrinsically enjoyable, and hence motivating. In this case, the theory was typically applied to making the tool/platform/tasks afford users a certain amount of autonomy, competence, and/or relatedness, such that they find using the BCT engaging, sufficiently challenging, and encouraging. In this case, the idea is that sustained motivation may be achieved by making users \textit{want} to do the target behaviour more, via the BCT (e.g., by gamifying the \textit{doing of} the task), and hence, as long as they volitionally choose to use the BCT for its intrinsic enjoyment, they will be engaging in the target behaviour. In this case, the SDT mini-theories of CET and BPNT were, mostly implicitly, applied as theoretical frameworks. 

The second is understanding BCTs as a tool for helping users achieve specific personal goals, where the aim is helping users reflect on, and internalise the value of changing their behaviour in some personally-desired way. By internalising the value of the behaviour change---prompting users to reflect on what they wish to achieve with the BCT, and bringing it into congruence with their other goals and values---the idea is that they may become more self-determined to do things that are instrumentally important to them, but that they do not necessarily find intrinsically motivating (e.g., \textit{wanting} to change their lifestyle/behaviours for finding it personally important and meaningful, rather than because they know they \textit{have to}). In this case, sustained behaviour change is achieved by enhancing the quality of a person's motivation for doing the target behaviour, such that they might eventually no longer even require external scaffolds like a BCT to feel motivated. In this case, the SDT mini-theory of OIT was---again, mostly implicitly---applied to help users move towards integrated regulation of the ultimate outcomes/goals they aim to reach through the BCT. 

Of the two approaches, the first was taken by the vast majority (10/15) of reviewed papers.
Whilst both have potential utility, we believe that the latter may be more effective/appropriate for promoting sustained motivation in many, if not most, contexts of BCTs, for reasons outlined below.

On the one hand, increasing a user's inherent enjoyment of a BCT (e.g., through gamification or social scaffolds) may well increase their willingness to use the intervention long-term, thereby helping them, in effect, perform the target behaviour more. However, this may be less effective or sustainable as a behaviour change approach than internalising the value of the behaviour \textit{per se}. Firstly, technologies like apps tend to have limited shelf-lives, or are subject to changes (e.g., many apps start to require paid subscriptions in order for the company to stay profitable). If a user's motivation entirely depends on their enjoyment of the BCT, but they are no longer able, or willing, to use it due to external factors beyond their control, it could mean that they stop engaging with the target behaviour entirely. 

Secondly, just because a BCT, like an exergame, motivates users to do some version of a target behaviour (e.g., jumping on a trampoline), it does not necessarily mean that engaging in the behaviour it encourages as much as possible is appropriate or sufficient for meeting their ultimate personal goal with the behaviour change (e.g., becoming fit or healthy, or losing weight). As Lerch \textit{et al.}'s \cite{lerch} findings suggest, there are sometimes valid, and even important, reasons why people may want to stop using a BCT like a fitness app, e.g., deciding to get a gym membership instead, or feeling sufficiently motivated to meet their exercise goals in other ways than the BCT supports. 
In a similar vein, Spiel et al. \cite{spiel2018fitter} critique fitness trackers for embedding the assumption that ``more steps equals more health'', regardless of other factors that may be required for a healthy lifestyle. One possible consequence of gamifying the doing of a `healthy' behaviour as such, without helping users understand the value and importance of performing the behaviour in the right way, at the right frequency, or adapting their approach based on their changing needs, is that people may start to ``hack'' the game for rewards rather than caring about whether or not the exercise was done correctly. This may include waiting for the timer to run out without exercising for the sake of getting points, or waving one's arm to increase steps in a wearable step counter. 

Moreover, as the path of least resistance, such shortcuts may make users lose interest in the app over time, as they find it too easy to avoid doing behaviours in more effortful, goal-directed ways. Similarly, users might start pursuing the proxy behaviour at the cost of their overall health (e.g., running too far and too frequently), as they are incentivised to maximise their engagement with the app, rather than being regulated by the requirements of meeting a specific personal goal. 
Finally, as mentioned earlier, intrinsic motivation is not something that is always present but merely increases or decreases, as several reviewed papers claimed. Instead, it is a specific type of motivation that individuals can have towards certain behaviours for certain reasons, which cannot necessarily be elicited for any target behaviour \cite{ryan2000intrinsic,ryan2017self}. Some behaviours are simply less inherently enjoyable than others, and even if one can make a target behaviour more enjoyable (e.g., with a BCT gamifying the process of stopping an addiction), this will only be effective as long as using the BCT is comparatively \textit{more} enjoyable than engaging in the behaviour it is trying to avoid (e.g., enjoying being supported in not drinking more than drinking), which is a tall order. Ryan and Deci \cite{ryan2000self} also highlight this risk: 
\begin{quote}
Yet, however promising the serious games and gamification movements sound, their promise has not always yielded desired results. Many educational, training-focused, and  health-related games are transparently extrinsic in their focus. Moreover, when such educational and training-focused games are expected to compete for players’ attention against actual games ... they will typically lose because the way they are designed leaves them less interesting and need satisfying. Similarly, simply adding game elements to a workplace or learning task will not typically be sufficient to enhance motivation, particularly if the activity is not already intrinsically interesting \cite[p.529-530]{ryan2017self}.
\end{quote}
Thus, we believe that applying OIT to help people internalise the value and importance of appropriately performing the target behaviour would be more conducive to self-determined, sustained motivation towards personal goals, than merely encouraging increased and sustained BCT use in line with CET or BPNT. As Ryan and Deci \cite{ryan2017self} explain:
\begin{quote}
Numerous studies in varied behavioral domains and using various assessment strategies indicate that more autonomous and self- concordant motivation is associated with greater behavioral persistence, as specified in OIT’s Proposition IV. Clearly, when people more fully internalize the value and importance of a behavior or domain, they are more likely to maintain relevant behaviors and beliefs than when they engage in such behaviors for more controlled reasons. One result of this is a higher probability of actually achieving the goals people pursue \cite[p.213]{ryan2017self}.
\end{quote}
We believe that this gap poses a great opportunity for 
BCT designers to utilise SDT towards meeting personal goals that require difficult or laborious tasks that users may not easily find intrinsically motivating (or at least not as strongly as compared to not doing it). 

One can imagine various probable reasons why promoting increased BCT engagement, rather than promoting people's independent pursuit of behaviour change, is the more common aim in HCI research. For one, companies typically benefit financially from increased and sustained user engagement with the BCT. If this happens to promote healthy behaviours and lifestyle choices in the users, it can seem like a win-win situation. It is also easier to measure success through people's BCT use statistics than if they were to stop using the BCT, and this use data can also be useful to learn more about human behaviour, as well as how to keep improving and adapting BCT design. Moreover, it may be that there is simply more of a precedent in using gamification or other UX-enhancement tactics in interface design than applying theoretical constructs in more OIT-consistent ways. As only five of the reviewed papers regarded user studies that endured longer than a single use, of which only three endured longer than a month, it is hard to know whether CET-consistent approaches actually succeed in effectively sustaining behaviour change as intended.

\subsection{\textbf{How SDT-related constructs were translated into design features}}

\subsubsection{Design features to support users' basic psychological needs}

Across the reviewed papers, we generated 50 themes representing design suggestions for supporting one of the \textit{basic psychological needs} posited by SDT: 11 for supporting `autonomy', 22 for `competence', and 17 for `relatedness'. These themes showed some overlap with those identified by Villalobos-Z\'{u}\~{n}iga and Cherubini's \cite{villa} review of features related to SDT in behaviour change apps: particularly, \textit{customisation} and (tailored/guided) \textit{goal-setting} for supporting autonomy; \textit{rewards}, \textit{self-monitoring} and \textit{activity}(\textit{/progress}) \textit{tracking} for supporting competence, and \textit{peer comparison} and \textit{group competition} to support relatedness. 

How exactly all the features were implemented varied between domains and technologies, as did the extent to which they convincingly seemed to support the BPN in a theory-consistent way. Similar to what Tyack and Mekler \cite{games} found in the use of SDT for gameplay, papers were sometimes ``frought with dubious (and often uncited) claims as to how individual game [BCT] elements relate to SDT concepts'' \cite[p.9]{games}: e.g., that supporting the three BPNs, in some way, will \textit{increase intrinsic motivation}. Even if features apparently associate with a given construct, e.g., relatedness (having a ``warm and caring'' interaction with a social robot), few evaluated its validity or justified it in terms of its relation to the broader theory or mini-theories. Even if pleasant, is not necessarily true that interaction with a robot would satisfy relatedness in the same way as connections to other people (e.g., feeling a part of a group, improving one's social status, making a loved one happy) might, or that it will effectively motivate the user for the target behaviour. Similarly, merely including a feature that gives users a level of control over some aspect of the interface, e.g., being able to choose between two options, does not necessarily mean that one's quality of motivation for the \textit{behaviour} the app supports will increase. This would all have to be evaluated with valid empirical measures, as we discuss in section 5.4. This shortcoming is not limited to HCI research, however. In their meta-analysis of SDT-based interventions, Gillison \textit{et al.} \cite{metagillison} identified similar limitations across disciplines that leveraged SDT for health behaviour interventions.\footnote{Gilllison \textit{et al.} argue that insights into the efficacy of interventions that leverage SDT is ``limited by poor specification of the intervention techniques employed (i.e., investigators may state that they provided an autonomy supportive environment without stating how they did so), and by a lack of information about the impact of specific techniques on the mediators of change proposed within SDT (e.g., need support and motivation); that is, it is often assumed techniques will have the hypothesised impact on mediators without this being explicitly tested.''\cite[p.111]{metagillison}.}

More than just implementing any one feature with a high degree of fidelity to the theory, designers should also consider interaction effects---whether any one feature will be sufficient for supporting the BPN in question, and whether supporting any one BPN will be as effective in a given context without supporting any others. Hekler \textit{et al.} suggested this as another potential pitfall of applying behavioural theories in HCI: ``By focusing on individual constructs rather than whole frameworks, however, an HCI researcher might inadvertently design a system based on constructs that do not work independently but only in tandem with other constructs''  \cite[p.3309]{Hekler}. This point is further unpacked by Villalobos-Z\'{u}\~{n}iga and Cherubini: 
\begin{quote}
We suggest three open questions: (i) It is still unclear whether providing support for only one, or two of the basic needs can yield positive effects on a user’s motivation. (ii) ... we do not know whether implementing multiple features that support the same BPN would actually increase the overall positive effect, or be detrimental towards supporting self-determined action towards the target activity. (iii) ...  It is not clear whether a particular combination of supports for the three basic needs would be better suited to help users with varying levels of intrinsic motivation (measured at the onset of the intervention) \cite[p.20]{villa}. 
\end{quote}
Moreover, it is important to consider how effects may evolve: whether apparently effective supports for any given need may wane or even have a negative effect over time.\footnote{Ryan and Deci emphasise the importance of follow-up studies, as ``SDT expects that careful use of external controls can produce behavior change in the short term ... But the more important and penetrating issue concerns the persistence of that behavior change ... when the controls were no longer in effect ... and it is here that the autonomy support and resulting internalization and integration of behavioral regulations are so crucial.'' \cite[p.207-208]{ryan2017self}.} We, therefore, echo Hekler \textit{et al.}'s suggestion that designers should treat any proposed design features for supporting SDT constructs as hypotheses that require additional testing, and that the gap between theory and a concrete design should be bridged for every new technology \cite[p.3310]{Hekler}, given potentially significant contextual differences. In the next section, we collate our findings of some of such contextual factors that may affect the suitability of certain design features for different BCTs. 

\subsubsection{Design features to support autonomous regulation}

Beyond supporting specific BPNs, we also identified design suggestions for facilitating an internal perceived locus of causality/autonomous regulation, such that users may internalise motivation for the target behaviour. 

Here we identified suggestions for creating space in which patients can be involved in meaningful goal-setting; helping users internalise content; allowing increasing users' awareness of their progress; providing tailored feedback at different stages of the process; by identifying specific behavioural variables of different users in the design stages; promoting interaction with other users; helping the user feel valued and appreciated; clearly communicate the BCT's value; and extending the frequency of satisfying or educational moments over time. 

As in the case of the BPNT supporting features above, the relative suitability of such suggestions will likely vary between domains and implementations. The most robust approach would be to apply a specific mini-theory as a theoretical framework (in this case, OIT) as a basis for understanding (a) the role that a given SDT construct plays in the theory, and how it may be most meaningfully operationalised in the given context. Moreover, validated SDT measures should be applied to evaluate the construct validity of a given operalisation in a given implementation.

\subsection{Gaps and limitations in existing work}

Similar to Tyack and Mekler's \cite{games} finding of the use of SDT in HCI games research, we found that the majority of papers engage with SDT in a limited manner, rarely mentioning the mini-theories that give SDT-related constructs their operational definitions. As such, the meaning and role of constructs were sometimes described in ways that were overly simplistic or not wholly theory-consistent. Drawing from Hekler \textit{et al.} \cite{Hekler}, 
we discuss these in terms of some typical risks of applying behavioural theories in technology design.


For one thing, only four papers used complete validated SDT-related scales to measure SDT-related constructs, while another two used shortened versions. However, most papers used no validated SDT-related scales or questionnaires. Instead, they did one of three things: either they used an understanding of the theory as an interpretative lens (e.g., creating their own scale or coding qualitative data in terms of the SDT constructs they seem to support), or they measured constructs of user experience/satisfaction unrelated to SDT, or did no user evaluation. Whilst using the theory as an interpretative lens for analysing data is a valid approach, and fits well within the tradition of theory-driven qualitative methods in HCI and the social sciences, there are methods to lower the risk of confirmation bias, which have not been followed in the reviewed work. One option suggested by Hekler \textit{et al.} \cite{Hekler} is having a preformulated coding manual that contains ``likely responses in user feedback that would indicate that the technology was having or not having a theoretically postulated effect'' \cite[p.3311-3312]{Hekler} Otherwise, it is too easy for researchers to subconsciously interpret anything that participants say as vaguely related to the constructs of SDT. This risk is even greater given the level of generality of SDT as a metatheory,\footnote{A metatheory has the highest level of generality of behavioural theories, and identifies ``broad `levels' of inter-related associations and factors of influence on a behavior of interest'' \cite[p.3308]{Hekler}} which means that there is a lot of ambiguity in how constructs may be operationalised, and translating it into design guidelines and features requires a great deal of conceptual work. 

This is one of the `common pitfalls' of applying a behavioural theory from psychology in HCI, as outlined by Hekler \textit{et al.} \cite{Hekler}. Another common pitfall is ``picking only some constructs from a theory and thus losing the potency of the full conceptual framework for designing a system'' \cite[p.3312]{Hekler}, which we also noticed in our review. That is, whilst all of the reviewed papers employed SDT-related constructs like extrinsic/intrinsic motivation and the three BPNs, few papers described these terms of the relevant mini-theories that describe them---also echoing Tyack and Mekler's \cite{games} finding. Related to this, we also found that a third common pitfall applied: ``using selective constructs from a theory but making claims that are related to the full theory'' \cite[p.3312]{Hekler}. There was a tendency in the reviewed papers to confuse or selectively combine constructs from two of its subtheories, CET and OIT. As mentioned, some papers described \textit{intrinsic motivation} as something that is always present and increases or decreases with better support of the BPNs, rather than a specific \textit{type} of motivation, as the theory maintains. Instead, the theory describes a process of internalisation/integration that occurs as the \textit{reasons} underlying an action become internalised and brought into congruence with one’s other values. A related second point is that supporting either autonomy, competence and relatedness in an interface---in some form or other---sometimes appeared to be taken as sufficient to catalyse intrinsic motivation. As Hekler \textit{et al.} \cite{Hekler} maintain, translating constructs from a psychological theory into design features---especially a metatheory with such a great degree of generality---requires a great deal of conceptual and formative work, as it leaves space for much ambiguity.  For example, autonomy does not necessarily just mean ``more choice, somewhere'', and what constitutes the right amount of choice and the right kind of things to choose from should be carefully explored with each new implementation, and in consideration of other contextual factors like the domain, differences in needs between individuals or groups, and for a given user over time. 

Despite SDT's central aim of supporting sustained and integrated forms of motivation, we found another limitation in that reviewed HCI papers tend to utilise it more for the sake of making BCTs engaging and satisfying (drawing from OIT and BPNT), rather than offering scaffolding, information and support for helping users internalise the target behaviour \textit{per se} (in line with OIT). Whilst the former may help motivate users to use the BCT in the short term, the fact that motivation depends so much on the intervention (and the encouragement/feedback/incentives it provides) means that its source is still primarily extrinsic, and hence their ability to sustain motivation for the target behaviour itself, independently of the intervention, is not supported. This approach lacks a meaningful provision of autonomy in not encouraging users to internalise/integrate the personal \textit{value} of the target behaviour. Hence, once the novelty effect or users' `intrinsic motivation' to engage with the BCT wanes, if the latter was even achieved, users may not have the same personal investment/motivation in continuing with either the BCT or the behaviour. For encouraging more sustained behaviour change, beyond BCT dependence, we suggest drawing from OIT as a potentially fruitful solution to explore in future work. 


Apart from choosing an appropriate theory, we now propose some general guidance for operationalising SDT constructs in interface design, drawing from Hekler \textit{et al.}'s \cite{Hekler} suggestions for applying behavioural theories in HCI. 

\subsection{Future opportunities: designing for sustained motivation}
Robustly applying SDT to the design of BCTs involves understanding the appropriate aims and metrics that relate to the theory, as well as the appropriate methods for translating its constructs into design features. We briefly consider each of these below.

\subsubsection{Appropriate aims}
A limitation we noticed in existing work, is that researchers tend to assume that the (implicit) ultimate aim of applying SDT to BCT design should be to get users to engage more with the technology, rather than leveraging the power of the theory for helping people internalise target behaviours to the point of becoming self-regulated. A particular value of SDT is its ability to help designers give people the tools and support they need to reach this point of self-sufficiency, instead of simply making the intervention mechanisms less controlling or more fun. For lasting results, it might be preferable not to limit tasks to the interaction with the technology, especially if the activity has the potential of being rewarding and valuable in itself (e.g., exercising or developing mindfulness). In such cases, the aim should be to encourage users to gain as much as possible autonomous control over their regulation such that they eventually do not need the intervention anymore. This may involve empowering them with information about how and why the target behaviour is important, and knowledge about how to help themselves (competence) as well as letting them become active participants in their self-regulation in their lives outside of the technology (relatedness). 
If gamification strategies are used, it may be helpful to find ways of gamifying self-sufficiency (i.e., gaining knowledge about the value and purpose of the target behaviour, and skills regarding performing it appropriately without guidance), not just the ``doing of'' a given target behaviour (e.g., assuming that \textit{more steps} equal \textit{better health} \cite{spiel2018fitter}). 

\subsubsection{Appropriate methods and metrics for translating constructs into design features }
Another limitation we noticed, was finding robust (and theory-consistent) ways to test and validate design features that aim to support users in aims such as the above (in accounting for relevant SDT-related constructs), avoiding common errors like confirmation bias and issues of construct validity. 

Although the designed purpose of most BCTs is to effectively motivate changes in user behaviour, the lack of resources in HCI for conducting large-scale (and long-term) randomised controlled trials (RCTs) means that this is rarely demonstrated \cite{Hekler}. Of our reviewed papers, only three studies lasted longer than a month, of which none were RCTs. For more robust studies, Hekler \textit{et al.} suggest a few suitable approaches for theory-driven study designs in HCI research. One approach is mediational/path and moderation analyses, which aim to account for variations in relevant variables like \textit{how} a BCT works (i.e., the mediating variables: which of its features drive behavioural change) and \textit{for whom} or \textit{under what circumstances} it works best (i.e., moderating variables like individual/group and situational differences). According to Helker \textit{et al.}, understanding key mediator variables within guiding theories like SDT can ``allow HCI researchers to both support these constructs in their designs and to assess them in their evaluations instead of solely relying on more distal outcomes such as behaviors'' \cite[p.3311]{Hekler}. Another suggested approach, mentioned earlier, is using theory to help evaluate qualitative data, e.g., testing whether a technology is operating according to the relevant theoretical mechanisms by formulating ``\textit{a priori} expectations of likely responses in user feedback that would indicate that the technology was having or not having a theoretically postulated effect'' \cite[p.3312]{Hekler}---see Grimes and Grinter \cite{grimes2007designing} for an example. This contrasts with the more deductive approach that was sometimes taken in our reviewed work, i.e., using thematic analysis to code participant observation/interview data in terms of the constructs they seem related to. 

Beyond validating causal links between specific features and the relevant constructs, researchers in the social sciences and HCI have developed and validated several measures for measuring SDT-related constructs like BPN satisfaction in different domains (see Peters \textit{et al.}\cite{Peters2018-ra} for an overview), which few of the reviewed work utilised. Here, it is also important to consider \textit{which} (combination of) scales are most appropriate for a given context (e.g., measures for evaluating motivation/BPN support in therapy differs from evaluating it in fitness), as well as \textit{when} (and how frequently) to evaluate the BCT's effectivity, as motivation levels/quality may fluctuate and wane over time (e.g., when the novelty effect wears off, or if a user fails to assimilate and integrate a regulation). This may require adapting supporting features or motivational approaches.

\section{Limitations and future work}
Our review was limited to publications in HCI venues, which may have excluded potentially relevant research in related fields (e.g., psychology, healthcare, education) that applied SDT to BCT design. We expect that the risks of validity and bias that Hekler \textit{et al.} \cite{Hekler} describe would be lower in social scientific studies in fields like psychology, where the focus is more on evaluating effect sizes with RCTs than on interface design. As our focus was more on the latter, and we wanted to survey the precedents being set in the HCI community, we limited our scope to records in the ACM Digital Library.

As our review included late-breaking reports and extended abstracts, it may be the case that some apparent limitations in understanding or applying SDT were due to constraints related to the paper format. We decided to still include those formats as this is still a relatively new/exploratory research area for HCI and we wanted to give a broad overview of all the initial approaches to applying SDT in this domain, as it may pave the way for future research avenues. 

Our suggestion for applying the SDT mini-theory of OIT to BCT design is theoretical and has not been explored in practice/with end-users. This is something we aim to look at in future work.

\section{Conclusion}
In our review of 15 HCI papers that apply SDT to the design of behaviour change technologies (BCTs), we described themes in how HCI researchers have translated core SDT constructs into design features, and highlighted some opportunities for future work aiming to further encourage sustained behaviour change. We identified some limitations in how the theory has been understood and applied (e.g., to enhance users' motivation to use a technology rather than supporting users in internalising the regulation towards becoming more self-determined). We also highlighted risks of confirmation bias and bad construct validity in cases where the theory was applied in an HCI context without taking appropriate cautionary measures in the study design. In doing so, we explore how the HCI community might unlock more of the theory’s potential and better support sustained motivation with BCTs. Our work contributes to discourse on designing wellbeing-supportive technologies, as well as the growing body of research in the design and evaluation of BCTs and for HCI researchers interested in utilising SDT.

\bibliographystyle{ACM-Reference-Format}

\begin{thebibliography}{87}


\ifx \showCODEN    \undefined \def \showCODEN     #1{\unskip}     \fi
\ifx \showDOI      \undefined \def \showDOI       #1{#1}\fi
\ifx \showISBNx    \undefined \def \showISBNx     #1{\unskip}     \fi
\ifx \showISBNxiii \undefined \def \showISBNxiii  #1{\unskip}     \fi
\ifx \showISSN     \undefined \def \showISSN      #1{\unskip}     \fi
\ifx \showLCCN     \undefined \def \showLCCN      #1{\unskip}     \fi
\ifx \shownote     \undefined \def \shownote      #1{#1}          \fi
\ifx \showarticletitle \undefined \def \showarticletitle #1{#1}   \fi
\ifx \showURL      \undefined \def \showURL       {\relax}        \fi
\providecommand\bibfield[2]{#2}
\providecommand\bibinfo[2]{#2}
\providecommand\natexlab[1]{#1}
\providecommand\showeprint[2][]{arXiv:#2}

\bibitem[Arksey and O'Malley(2005)]%
        {arksey2005scoping}
\bibfield{author}{\bibinfo{person}{Hilary Arksey} {and} \bibinfo{person}{Lisa
  O'Malley}.} \bibinfo{year}{2005}\natexlab{}.
\newblock \showarticletitle{Scoping studies: towards a methodological
  framework}.
\newblock \bibinfo{journal}{\emph{International journal of social research
  methodology}} \bibinfo{volume}{8}, \bibinfo{number}{1}
  (\bibinfo{year}{2005}), \bibinfo{pages}{19--32}.
\newblock


\bibitem[Aufheimer et~al\mbox{.}(2023)]%
        {Aufheimer}
\bibfield{author}{\bibinfo{person}{Maria Aufheimer}, \bibinfo{person}{Kathrin
  Gerling}, \bibinfo{person}{T.C.~Nicholas Graham}, \bibinfo{person}{Mari
  Naaris}, \bibinfo{person}{Marco~J. Konings}, \bibinfo{person}{Elegast
  Monbaliu}, \bibinfo{person}{Hans Hallez}, {and} \bibinfo{person}{Els
  Ortibus}.} \bibinfo{year}{2023}\natexlab{}.
\newblock \showarticletitle{An Examination of Motivation in Physical Therapy
  Through the Lens of Self-Determination Theory: Implications for Game Design}.
  In \bibinfo{booktitle}{\emph{Proceedings of the 2023 CHI Conference on Human
  Factors in Computing Systems}} (Hamburg, Germany) \emph{(\bibinfo{series}{CHI
  '23})}. \bibinfo{publisher}{Association for Computing Machinery},
  \bibinfo{address}{New York, NY, USA}, Article \bibinfo{articleno}{725},
  \bibinfo{numpages}{16}~pages.
\newblock
\showISBNx{9781450394215}
\urldef\tempurl%
\url{https://doi.org/10.1145/3544548.3581171}
\showDOI{\tempurl}


\bibitem[Ballou et~al\mbox{.}(2022a)]%
        {Ballou2022SDTinHCI}
\bibfield{author}{\bibinfo{person}{Nick Ballou}, \bibinfo{person}{Sebastian
  Deterding}, \bibinfo{person}{April Tyack}, \bibinfo{person}{Elisa~D Mekler},
  \bibinfo{person}{Rafael~A Calvo}, \bibinfo{person}{Dorian Peters},
  \bibinfo{person}{Gabriela Villalobos-Z\'{u}\~{n}iga}, {and}
  \bibinfo{person}{Selen Turkay}.} \bibinfo{year}{2022}\natexlab{a}.
\newblock \showarticletitle{Self-Determination Theory in HCI: Shaping a
  Research Agenda}. In \bibinfo{booktitle}{\emph{Extended Abstracts of the 2022
  CHI Conference on Human Factors in Computing Systems}} (New Orleans, LA, USA)
  \emph{(\bibinfo{series}{CHI EA '22})}. \bibinfo{publisher}{Association for
  Computing Machinery}, \bibinfo{address}{New York, NY, USA}, Article
  \bibinfo{articleno}{113}, \bibinfo{numpages}{6}~pages.
\newblock
\showISBNx{9781450391566}
\urldef\tempurl%
\url{https://doi.org/10.1145/3491101.3503702}
\showDOI{\tempurl}


\bibitem[Ballou et~al\mbox{.}(2022b)]%
        {Shaping}
\bibfield{author}{\bibinfo{person}{Nick Ballou}, \bibinfo{person}{Sebastian
  Deterding}, \bibinfo{person}{April Tyack}, \bibinfo{person}{Elisa~D Mekler},
  \bibinfo{person}{Rafael~A Calvo}, \bibinfo{person}{Dorian Peters},
  \bibinfo{person}{Gabriela Villalobos-Z\'{u}\~{n}iga}, {and}
  \bibinfo{person}{Selen Turkay}.} \bibinfo{year}{2022}\natexlab{b}.
\newblock \showarticletitle{Self-Determination Theory in HCI: Shaping a
  Research Agenda}. In \bibinfo{booktitle}{\emph{Extended Abstracts of the 2022
  CHI Conference on Human Factors in Computing Systems}} (New Orleans, LA, USA)
  \emph{(\bibinfo{series}{CHI EA '22})}. \bibinfo{publisher}{Association for
  Computing Machinery}, \bibinfo{address}{New York, NY, USA}, Article
  \bibinfo{articleno}{113}, \bibinfo{numpages}{6}~pages.
\newblock
\showISBNx{9781450391566}
\urldef\tempurl%
\url{https://doi.org/10.1145/3491101.3503702}
\showDOI{\tempurl}


\bibitem[Bargas-Avila and Hornb\ae{}k(2011)]%
        {sys1}
\bibfield{author}{\bibinfo{person}{Javier~A. Bargas-Avila} {and}
  \bibinfo{person}{Kasper Hornb\ae{}k}.} \bibinfo{year}{2011}\natexlab{}.
\newblock \showarticletitle{Old wine in new bottles or novel challenges: a
  critical analysis of empirical studies of user experience}. In
  \bibinfo{booktitle}{\emph{Proceedings of the SIGCHI Conference on Human
  Factors in Computing Systems}} (<conf-loc>, <city>Vancouver</city>,
  <state>BC</state>, <country>Canada</country>, </conf-loc>)
  \emph{(\bibinfo{series}{CHI '11})}. \bibinfo{publisher}{Association for
  Computing Machinery}, \bibinfo{address}{New York, NY, USA},
  \bibinfo{pages}{2689–2698}.
\newblock
\showISBNx{9781450302289}
\urldef\tempurl%
\url{https://doi.org/10.1145/1978942.1979336}
\showDOI{\tempurl}


\bibitem[Behnke et~al\mbox{.}(2016)]%
        {7}
\bibfield{author}{\bibinfo{person}{Kara~Alexandra Behnke},
  \bibinfo{person}{Brittany~Ann Kos}, {and} \bibinfo{person}{John~K. Bennett}.}
  \bibinfo{year}{2016}\natexlab{}.
\newblock \showarticletitle{Computer Science Principles: Impacting Student
  Motivation and Learning Within and Beyond the Classroom}. In
  \bibinfo{booktitle}{\emph{Proceedings of the 2016 ACM Conference on
  International Computing Education Research}} (Melbourne, VIC, Australia)
  \emph{(\bibinfo{series}{ICER '16})}. \bibinfo{publisher}{Association for
  Computing Machinery}, \bibinfo{address}{New York, NY, USA},
  \bibinfo{pages}{171–180}.
\newblock
\showISBNx{9781450344494}
\urldef\tempurl%
\url{https://doi.org/10.1145/2960310.2960336}
\showDOI{\tempurl}


\bibitem[Bomfim and Wallace(2018)]%
        {Bomfim}
\bibfield{author}{\bibinfo{person}{Marcela C.~C. Bomfim} {and}
  \bibinfo{person}{James~R. Wallace}.} \bibinfo{year}{2018}\natexlab{}.
\newblock \showarticletitle{Pirate Bri's Grocery Adventure: Teaching Food
  Literacy through Shopping}. In \bibinfo{booktitle}{\emph{Extended Abstracts
  of the 2018 CHI Conference on Human Factors in Computing Systems}} (Montreal
  QC, Canada) \emph{(\bibinfo{series}{CHI EA '18})}.
  \bibinfo{publisher}{Association for Computing Machinery},
  \bibinfo{address}{New York, NY, USA}, \bibinfo{pages}{1–6}.
\newblock
\showISBNx{9781450356213}
\urldef\tempurl%
\url{https://doi.org/10.1145/3170427.3188496}
\showDOI{\tempurl}


\bibitem[Braun and Clarke(2019)]%
        {braun2019reflecting}
\bibfield{author}{\bibinfo{person}{Virginia Braun} {and}
  \bibinfo{person}{Victoria Clarke}.} \bibinfo{year}{2019}\natexlab{}.
\newblock \showarticletitle{Reflecting on reflexive thematic analysis}.
\newblock \bibinfo{journal}{\emph{Qualitative research in sport, exercise and
  health}} \bibinfo{volume}{11}, \bibinfo{number}{4} (\bibinfo{year}{2019}),
  \bibinfo{pages}{589--597}.
\newblock


\bibitem[Br\"{u}hlmann et~al\mbox{.}(2018)]%
        {Brhlmann2018UMI}
\bibfield{author}{\bibinfo{person}{Florian Br\"{u}hlmann},
  \bibinfo{person}{Beat Vollenwyder}, \bibinfo{person}{Klaus Opwis}, {and}
  \bibinfo{person}{Elisa~D. Mekler}.} \bibinfo{year}{2018}\natexlab{}.
\newblock \showarticletitle{Measuring the “Why” of Interaction: Development
  and Validation of the User Motivation Inventory (UMI)}. In
  \bibinfo{booktitle}{\emph{Proceedings of the 2018 CHI Conference on Human
  Factors in Computing Systems}} (<conf-loc>, <city>Montreal QC</city>,
  <country>Canada</country>, </conf-loc>) \emph{(\bibinfo{series}{CHI '18})}.
  \bibinfo{publisher}{Association for Computing Machinery},
  \bibinfo{address}{New York, NY, USA}, \bibinfo{pages}{1–13}.
\newblock
\showISBNx{9781450356206}
\urldef\tempurl%
\url{https://doi.org/10.1145/3173574.3173680}
\showDOI{\tempurl}


\bibitem[{Center for Self-Determination Theory}(2022)]%
        {centerSDTQuestionnaires}
\bibfield{author}{\bibinfo{person}{{Center for Self-Determination Theory}}.}
  \bibinfo{year}{2022}\natexlab{}.
\newblock \bibinfo{title}{Metrics \& {{Methods}}: {{Questionnaires}}}.
\newblock
  \bibinfo{howpublished}{\url{https://selfdeterminationtheory.org/questionnaires/}}.
\newblock
\newblock
\shownote{Accessed: 2022-02-23}.


\bibitem[Chaudhry et~al\mbox{.}(2022)]%
        {Chaudhry}
\bibfield{author}{\bibinfo{person}{Beenish~Moalla Chaudhry},
  \bibinfo{person}{Dipanwita Dasgupta}, {and} \bibinfo{person}{Nitesh Chawla}.}
  \bibinfo{year}{2022}\natexlab{}.
\newblock \showarticletitle{Formative Evaluation of a Tablet Application to
  Support Goal-Oriented Care in Community-Dwelling Older Adults}.
\newblock \bibinfo{journal}{\emph{Proc. ACM Hum.-Comput. Interact.}}
  \bibinfo{volume}{6}, \bibinfo{number}{MHCI}, Article \bibinfo{articleno}{208}
  (\bibinfo{date}{sep} \bibinfo{year}{2022}), \bibinfo{numpages}{21}~pages.
\newblock
\urldef\tempurl%
\url{https://doi.org/10.1145/3546743}
\showDOI{\tempurl}


\bibitem[Chen et~al\mbox{.}(2014)]%
        {Chen2014BPNS}
\bibfield{author}{\bibinfo{person}{Beiwen Chen}, \bibinfo{person}{Maarten
  Vansteenkiste}, \bibinfo{person}{Wim Beyers}, \bibinfo{person}{Liesbet
  Boone}, \bibinfo{person}{Edward~L. Deci}, \bibinfo{person}{Jolene~Van der
  Kaap-Deeder}, \bibinfo{person}{Bart Duriez}, \bibinfo{person}{Willy Lens},
  \bibinfo{person}{Lennia Matos}, \bibinfo{person}{Athanasios Mouratidis},
  \bibinfo{person}{Richard~M. Ryan}, \bibinfo{person}{Kennon~M. Sheldon},
  \bibinfo{person}{Bart Soenens}, \bibinfo{person}{Stijn~Van Petegem}, {and}
  \bibinfo{person}{Joke Verstuyf}.} \bibinfo{year}{2014}\natexlab{}.
\newblock \showarticletitle{Basic psychological need satisfaction, need
  frustration, and need strength across four cultures}.
\newblock \bibinfo{journal}{\emph{Motivation and Emotion}}
  \bibinfo{volume}{39}, \bibinfo{number}{2} (\bibinfo{date}{Nov.}
  \bibinfo{year}{2014}), \bibinfo{pages}{216--236}.
\newblock
\urldef\tempurl%
\url{https://doi.org/10.1007/s11031-014-9450-1}
\showDOI{\tempurl}


\bibitem[Clarke et~al\mbox{.}(2015)]%
        {clarke2015thematic}
\bibfield{author}{\bibinfo{person}{Victoria Clarke}, \bibinfo{person}{Virginia
  Braun}, {and} \bibinfo{person}{Nikki Hayfield}.}
  \bibinfo{year}{2015}\natexlab{}.
\newblock \showarticletitle{Thematic analysis}.
\newblock \bibinfo{journal}{\emph{Qualitative psychology: A practical guide to
  research methods}} \bibinfo{volume}{222}, \bibinfo{number}{2015}
  (\bibinfo{year}{2015}), \bibinfo{pages}{248}.
\newblock


\bibitem[Consolvo et~al\mbox{.}(2008)]%
        {consolvo2008activity}
\bibfield{author}{\bibinfo{person}{Sunny Consolvo}, \bibinfo{person}{David~W.
  McDonald}, \bibinfo{person}{Tammy Toscos}, \bibinfo{person}{Mike~Y. Chen},
  \bibinfo{person}{Jon Froehlich}, \bibinfo{person}{Beverly Harrison},
  \bibinfo{person}{Predrag Klasnja}, \bibinfo{person}{Anthony LaMarca},
  \bibinfo{person}{Louis LeGrand}, \bibinfo{person}{Ryan Libby},
  \bibinfo{person}{Ian Smith}, {and} \bibinfo{person}{James~A. Landay}.}
  \bibinfo{year}{2008}\natexlab{}.
\newblock \showarticletitle{Activity sensing in the wild: a field trial of
  ubifit garden}. In \bibinfo{booktitle}{\emph{Proceedings of the SIGCHI
  Conference on Human Factors in Computing Systems}} (Florence, Italy)
  \emph{(\bibinfo{series}{CHI '08})}. \bibinfo{publisher}{Association for
  Computing Machinery}, \bibinfo{address}{New York, NY, USA},
  \bibinfo{pages}{1797–1806}.
\newblock
\showISBNx{9781605580111}
\urldef\tempurl%
\url{https://doi.org/10.1145/1357054.1357335}
\showDOI{\tempurl}


\bibitem[Cox et~al\mbox{.}(2016)]%
        {Cox2016DesignFrictions}
\bibfield{author}{\bibinfo{person}{Anna~L Cox}, \bibinfo{person}{Sandy J~J
  Gould}, \bibinfo{person}{Marta~E Cecchinato}, \bibinfo{person}{Ioanna
  Iacovides}, {and} \bibinfo{person}{Ian Renfree}.}
  \bibinfo{year}{2016}\natexlab{}.
\newblock \showarticletitle{Design Frictions for Mindful Interactions: The Case
  for Microboundaries}. In \bibinfo{booktitle}{\emph{Proceedings of the 2016
  {CHI} Conference Extended Abstracts on Human Factors in Computing Systems}}
  (San Jose, California, USA) \emph{(\bibinfo{series}{CHI EA '16})}.
  \bibinfo{publisher}{Association for Computing Machinery},
  \bibinfo{address}{New York, NY, USA}, \bibinfo{pages}{1389--1397}.
\newblock
\showISBNx{9781450340823}
\urldef\tempurl%
\url{https://doi.org/10.1145/2851581.2892410}
\showDOI{\tempurl}


\bibitem[Curry(2021)]%
        {app}
\bibfield{author}{\bibinfo{person}{David Curry}.}
  \bibinfo{year}{2021}\natexlab{}.
\newblock \bibinfo{booktitle}{\emph{App Data Report 2023}}.
\newblock \bibinfo{type}{{T}echnical {R}eport}. \bibinfo{institution}{Business
  of Apps}.
\newblock
\urldef\tempurl%
\url{https://www.businessofapps.com/data/report-app-data/}
\showURL{%
\tempurl}


\bibitem[Davis et~al\mbox{.}(1992)]%
        {davis1992extrinsic}
\bibfield{author}{\bibinfo{person}{Fred~D Davis}, \bibinfo{person}{Richard~P
  Bagozzi}, {and} \bibinfo{person}{Paul~R Warshaw}.}
  \bibinfo{year}{1992}\natexlab{}.
\newblock \showarticletitle{Extrinsic and intrinsic motivation to use computers
  in the workplace 1}.
\newblock \bibinfo{journal}{\emph{Journal of applied social psychology}}
  \bibinfo{volume}{22}, \bibinfo{number}{14} (\bibinfo{year}{1992}),
  \bibinfo{pages}{1111--1132}.
\newblock


\bibitem[Davis et~al\mbox{.}(2016)]%
        {davis2016motivating}
\bibfield{author}{\bibinfo{person}{William~E Davis},
  \bibinfo{person}{Nicholas~J Kelley}, \bibinfo{person}{Jinhyung Kim},
  \bibinfo{person}{David Tang}, {and} \bibinfo{person}{Joshua~A Hicks}.}
  \bibinfo{year}{2016}\natexlab{}.
\newblock \showarticletitle{Motivating the academic mind: High-level construal
  of academic goals enhances goal meaningfulness, motivation, and
  self-concordance}.
\newblock \bibinfo{journal}{\emph{Motivation and Emotion}}
  \bibinfo{volume}{40} (\bibinfo{year}{2016}), \bibinfo{pages}{193--202}.
\newblock


\bibitem[Deci and Ryan(1985)]%
        {deci1985intrinsic}
\bibfield{author}{\bibinfo{person}{Edward~L Deci} {and} \bibinfo{person}{RM
  Ryan}.} \bibinfo{year}{1985}\natexlab{}.
\newblock \showarticletitle{Intrinsic Motivation and Self-Determination in
  Human Behavior}.
\newblock In \bibinfo{booktitle}{\emph{Springer Science \& Business Media}}.
  \bibinfo{publisher}{Springer}, \bibinfo{address}{Berlin}.
\newblock
\urldef\tempurl%
\url{https://doi.org/10.1007/978-1-4899-2271-7}
\showURL{%
\tempurl}


\bibitem[Edmunds et~al\mbox{.}(2008)]%
        {edmunds2008testing}
\bibfield{author}{\bibinfo{person}{Jemma Edmunds}, \bibinfo{person}{Nikos
  Ntoumanis}, {and} \bibinfo{person}{Joan~L Duda}.}
  \bibinfo{year}{2008}\natexlab{}.
\newblock \showarticletitle{Testing a self-determination theory-based teaching
  style intervention in the exercise domain}.
\newblock \bibinfo{journal}{\emph{European journal of social psychology}}
  \bibinfo{volume}{38}, \bibinfo{number}{2} (\bibinfo{year}{2008}),
  \bibinfo{pages}{375--388}.
\newblock


\bibitem[Erskine et~al\mbox{.}(2010)]%
        {erskine2010suppress}
\bibfield{author}{\bibinfo{person}{James~AK Erskine}, \bibinfo{person}{George~J
  Georgiou}, {and} \bibinfo{person}{Lia Kvavilashvili}.}
  \bibinfo{year}{2010}\natexlab{}.
\newblock \showarticletitle{I suppress, therefore I smoke: Effects of thought
  suppression on smoking behavior}.
\newblock \bibinfo{journal}{\emph{Psychological science}} \bibinfo{volume}{21},
  \bibinfo{number}{9} (\bibinfo{year}{2010}), \bibinfo{pages}{1225--1230}.
\newblock


\bibitem[Ferron and Massa(2013)]%
        {Ferron}
\bibfield{author}{\bibinfo{person}{Michela Ferron} {and} \bibinfo{person}{Paolo
  Massa}.} \bibinfo{year}{2013}\natexlab{}.
\newblock \showarticletitle{Transtheoretical Model for Designing Technologies
  Supporting an Active Lifestyle}. In \bibinfo{booktitle}{\emph{Proceedings of
  the Biannual Conference of the Italian Chapter of SIGCHI}} (Trento, Italy)
  \emph{(\bibinfo{series}{CHItaly '13})}. \bibinfo{publisher}{Association for
  Computing Machinery}, \bibinfo{address}{New York, NY, USA}, Article
  \bibinfo{articleno}{7}, \bibinfo{numpages}{8}~pages.
\newblock
\showISBNx{9781450320610}
\urldef\tempurl%
\url{https://doi.org/10.1145/2499149.2499158}
\showDOI{\tempurl}


\bibitem[Fitzsimons and Lehmann(2004)]%
        {Fitzsimons2004-qx}
\bibfield{author}{\bibinfo{person}{Gavan~J Fitzsimons} {and}
  \bibinfo{person}{Donald~R Lehmann}.} \bibinfo{year}{2004}\natexlab{}.
\newblock \showarticletitle{Reactance to Recommendations: When Unsolicited
  Advice Yields Contrary Responses}.
\newblock \bibinfo{journal}{\emph{Marketing Science}} \bibinfo{volume}{23},
  \bibinfo{number}{1} (\bibinfo{date}{Feb.} \bibinfo{year}{2004}),
  \bibinfo{pages}{82--94}.
\newblock
\showISSN{0732-2399}
\urldef\tempurl%
\url{https://doi.org/10.1287/mksc.1030.0033}
\showDOI{\tempurl}


\bibitem[Ford et~al\mbox{.}(2012)]%
        {Fordd}
\bibfield{author}{\bibinfo{person}{Matthew Ford}, \bibinfo{person}{Peta Wyeth},
  {and} \bibinfo{person}{Daniel Johnson}.} \bibinfo{year}{2012}\natexlab{}.
\newblock \showarticletitle{Self-Determination Theory as Applied to the Design
  of a Software Learning System Using Whole-Body Controls}. In
  \bibinfo{booktitle}{\emph{Proceedings of the 24th Australian Computer-Human
  Interaction Conference}} (Melbourne, Australia) \emph{(\bibinfo{series}{OzCHI
  '12})}. \bibinfo{publisher}{Association for Computing Machinery},
  \bibinfo{address}{New York, NY, USA}, \bibinfo{pages}{146–149}.
\newblock
\showISBNx{9781450314381}
\urldef\tempurl%
\url{https://doi.org/10.1145/2414536.2414562}
\showDOI{\tempurl}


\bibitem[Fortier et~al\mbox{.}(2012)]%
        {Fortier_Duda_Guerin_Teixeira_2012}
\bibfield{author}{\bibinfo{person}{Michelle~S. Fortier},
  \bibinfo{person}{Joan~L. Duda}, \bibinfo{person}{Eva Guerin}, {and}
  \bibinfo{person}{Pedro~J. Teixeira}.} \bibinfo{year}{2012}\natexlab{}.
\newblock \showarticletitle{Promoting physical activity: development and
  testing of self-determination theory-based interventions}.
\newblock \bibinfo{journal}{\emph{International Journal of Behavioral Nutrition
  and Physical Activity}} \bibinfo{volume}{9}, \bibinfo{number}{1}
  (\bibinfo{date}{March} \bibinfo{year}{2012}), \bibinfo{pages}{20}.
\newblock
\showISSN{1479-5868}
\urldef\tempurl%
\url{https://doi.org/10.1186/1479-5868-9-20}
\showDOI{\tempurl}


\bibitem[Gillison et~al\mbox{.}(2019)]%
        {metagillison}
\bibfield{author}{\bibinfo{person}{Fiona Gillison}, \bibinfo{person}{Peter
  Rouse}, \bibinfo{person}{Martyn Standage}, \bibinfo{person}{Simon~J. Sebire},
  {and} \bibinfo{person}{Richard~M. Ryan}.} \bibinfo{year}{2019}\natexlab{}.
\newblock \showarticletitle{A meta-analysis of techniques to promote motivation
  for health behaviour change from a self-determination theory perspective}.
\newblock \bibinfo{journal}{\emph{Health Psychology Review}}
  \bibinfo{volume}{13}, \bibinfo{number}{1} (\bibinfo{year}{2019}),
  \bibinfo{pages}{110--130}.
\newblock
\urldef\tempurl%
\url{https://doi.org/10.1080/17437199.2018.1534071}
\showDOI{\tempurl}


\bibitem[Grau et~al\mbox{.}(2018)]%
        {26}
\bibfield{author}{\bibinfo{person}{Paul Grau}, \bibinfo{person}{Babak Naderi},
  {and} \bibinfo{person}{Juho Kim}.} \bibinfo{year}{2018}\natexlab{}.
\newblock \showarticletitle{Personalized Motivation-Supportive Messages for
  Increasing Participation in Crowd-Civic Systems}.
\newblock \bibinfo{journal}{\emph{Proc. ACM Hum.-Comput. Interact.}}
  \bibinfo{volume}{2}, \bibinfo{number}{CSCW}, Article \bibinfo{articleno}{60}
  (\bibinfo{date}{nov} \bibinfo{year}{2018}), \bibinfo{numpages}{22}~pages.
\newblock
\urldef\tempurl%
\url{https://doi.org/10.1145/3274329}
\showDOI{\tempurl}


\bibitem[Grimes and Grinter(2007)]%
        {grimes2007designing}
\bibfield{author}{\bibinfo{person}{Andrea Grimes} {and}
  \bibinfo{person}{Rebecca~E Grinter}.} \bibinfo{year}{2007}\natexlab{}.
\newblock \showarticletitle{Designing persuasion: Health technology for
  low-income African American communities}. In
  \bibinfo{booktitle}{\emph{Persuasive Technology: Second International
  Conference on Persuasive Technology}}. Springer,
  \bibinfo{publisher}{PERSUASIVE 2007}, \bibinfo{address}{Palo Alto, CA, USA},
  \bibinfo{pages}{24--35}.
\newblock


\bibitem[GVR(2021)]%
        {report}
\bibfield{author}{\bibinfo{person}{GVR}.} \bibinfo{year}{2021}\natexlab{}.
\newblock \bibinfo{booktitle}{\emph{Digital Health Market Size, Share and
  Trends Analysis Report By Technology (Healthcare Analytics, mHealth,
  Tele-healthcare, Digital Health Systems), By Component (Software, Hardware,
  Services), By Region, And Segment Forecasts, 2023 - 2030}}.
\newblock \bibinfo{type}{{T}echnical {R}eport} GVR-2-68038-886-2.
  \bibinfo{institution}{Grand View Research}.
\newblock


\bibitem[Harrison et~al\mbox{.}(2015)]%
        {harrison2015activity}
\bibfield{author}{\bibinfo{person}{Daniel Harrison}, \bibinfo{person}{Paul
  Marshall}, \bibinfo{person}{Nadia Bianchi-Berthouze}, {and}
  \bibinfo{person}{Jon Bird}.} \bibinfo{year}{2015}\natexlab{}.
\newblock \showarticletitle{Activity tracking: barriers, workarounds and
  customisation}. In \bibinfo{booktitle}{\emph{Proceedings of the 2015 ACM
  International Joint Conference on Pervasive and Ubiquitous Computing}}
  (Osaka, Japan) \emph{(\bibinfo{series}{UbiComp '15})}.
  \bibinfo{publisher}{Association for Computing Machinery},
  \bibinfo{address}{New York, NY, USA}, \bibinfo{pages}{617–621}.
\newblock
\showISBNx{9781450335744}
\urldef\tempurl%
\url{https://doi.org/10.1145/2750858.2805832}
\showDOI{\tempurl}


\bibitem[Hekler et~al\mbox{.}(2013)]%
        {Hekler}
\bibfield{author}{\bibinfo{person}{Eric~B. Hekler}, \bibinfo{person}{Predrag
  Klasnja}, \bibinfo{person}{Jon~E. Froehlich}, {and}
  \bibinfo{person}{Matthew~P. Buman}.} \bibinfo{year}{2013}\natexlab{}.
\newblock \showarticletitle{Mind the Theoretical Gap: Interpreting, Using, and
  Developing Behavioral Theory in HCI Research}. In
  \bibinfo{booktitle}{\emph{Proceedings of the SIGCHI Conference on Human
  Factors in Computing Systems}} (Paris, France) \emph{(\bibinfo{series}{CHI
  '13})}. \bibinfo{publisher}{Association for Computing Machinery},
  \bibinfo{address}{New York, NY, USA}, \bibinfo{pages}{3307–3316}.
\newblock
\showISBNx{9781450318990}
\urldef\tempurl%
\url{https://doi.org/10.1145/2470654.2466452}
\showDOI{\tempurl}


\bibitem[Honary et~al\mbox{.}(2019)]%
        {honary2019understanding}
\bibfield{author}{\bibinfo{person}{Mahsa Honary}, \bibinfo{person}{Beth~T
  Bell}, \bibinfo{person}{Sarah Clinch}, \bibinfo{person}{Sarah~E Wild},
  \bibinfo{person}{Roisin McNaney}, {et~al\mbox{.}}}
  \bibinfo{year}{2019}\natexlab{}.
\newblock \showarticletitle{Understanding the role of healthy eating and
  fitness mobile apps in the formation of maladaptive eating and exercise
  behaviors in young people}.
\newblock \bibinfo{journal}{\emph{JMIR mHealth and uHealth}}
  \bibinfo{volume}{7}, \bibinfo{number}{6} (\bibinfo{year}{2019}),
  \bibinfo{pages}{e14239}.
\newblock


\bibitem[Isaac(2021)]%
        {Isaac2021-ky}
\bibfield{author}{\bibinfo{person}{Sara Isaac}.}
  \bibinfo{year}{2021}\natexlab{}.
\newblock \bibinfo{title}{Why You Hate Being Told What to Do --- Even When
  You're Talking to Yourself}.
\newblock
  \bibinfo{howpublished}{\url{https://funeasypopular.com/why-you-hate-being-told-what-to-do-even-when-youre-talking-to-yourself/}}.
\newblock
\urldef\tempurl%
\url{https://funeasypopular.com/why-you-hate-being-told-what-to-do-even-when-youre-talking-to-yourself/}
\showURL{%
\tempurl}
\newblock
\shownote{Accessed: 2022-2-22}.


\bibitem[Jansen et~al\mbox{.}(2017)]%
        {Jansen}
\bibfield{author}{\bibinfo{person}{Arne Jansen}, \bibinfo{person}{Maarten
  Van~Mechelen}, {and} \bibinfo{person}{Karin Slegers}.}
  \bibinfo{year}{2017}\natexlab{}.
\newblock \showarticletitle{Personas and Behavioral Theories: A Case Study
  Using Self-Determination Theory to Construct Overweight Personas}. In
  \bibinfo{booktitle}{\emph{Proceedings of the 2017 CHI Conference on Human
  Factors in Computing Systems}} (Denver, Colorado, USA)
  \emph{(\bibinfo{series}{CHI '17})}. \bibinfo{publisher}{Association for
  Computing Machinery}, \bibinfo{address}{New York, NY, USA},
  \bibinfo{pages}{2127–2136}.
\newblock
\showISBNx{9781450346559}
\urldef\tempurl%
\url{https://doi.org/10.1145/3025453.3026003}
\showDOI{\tempurl}


\bibitem[Kim et~al\mbox{.}(2019)]%
        {Kim2019Goalkeeper}
\bibfield{author}{\bibinfo{person}{Jaejeung Kim}, \bibinfo{person}{Hayoung
  Jung}, \bibinfo{person}{Minsam Ko}, {and} \bibinfo{person}{Uichin Lee}.}
  \bibinfo{year}{2019}\natexlab{}.
\newblock \showarticletitle{{GoalKeeper}: Exploring Interaction Lockout
  Mechanisms for Regulating Smartphone Use}.
\newblock \bibinfo{journal}{\emph{Proc. ACM Interact. Mob. Wearable Ubiquitous
  Technol.}} \bibinfo{volume}{3}, \bibinfo{number}{1} (\bibinfo{date}{March}
  \bibinfo{year}{2019}), \bibinfo{pages}{29}.
\newblock
\urldef\tempurl%
\url{https://doi.org/10.1145/3314403}
\showDOI{\tempurl}


\bibitem[Korn and Tietz(2017)]%
        {36}
\bibfield{author}{\bibinfo{person}{Oliver Korn} {and} \bibinfo{person}{Stefan
  Tietz}.} \bibinfo{year}{2017}\natexlab{}.
\newblock \showarticletitle{Strategies for Playful Design When Gamifying
  Rehabilitation: A Study on User Experience}. In
  \bibinfo{booktitle}{\emph{Proceedings of the 10th International Conference on
  PErvasive Technologies Related to Assistive Environments}} (Island of Rhodes,
  Greece) \emph{(\bibinfo{series}{PETRA '17})}. \bibinfo{publisher}{Association
  for Computing Machinery}, \bibinfo{address}{New York, NY, USA},
  \bibinfo{pages}{209–214}.
\newblock
\showISBNx{9781450352277}
\urldef\tempurl%
\url{https://doi.org/10.1145/3056540.3056550}
\showDOI{\tempurl}


\bibitem[Kovacs et~al\mbox{.}(2021)]%
        {Kovacs2021NotNowAskAgainLater}
\bibfield{author}{\bibinfo{person}{Geza Kovacs}, \bibinfo{person}{Zhengxuan
  Wu}, {and} \bibinfo{person}{Michael~S. Bernstein}.}
  \bibinfo{year}{2021}\natexlab{}.
\newblock \showarticletitle{Not Now, Ask Later: Users Weaken Their Behavior
  Change Regimen Over Time, But Expect To Re-Strengthen It Imminently}. In
  \bibinfo{booktitle}{\emph{Proceedings of the 2021 CHI Conference on Human
  Factors in Computing Systems}} (Yokohama, Japan) \emph{(\bibinfo{series}{CHI
  '21})}. \bibinfo{publisher}{Association for Computing Machinery},
  \bibinfo{address}{New York, NY, USA}, Article \bibinfo{articleno}{229},
  \bibinfo{numpages}{14}~pages.
\newblock
\showISBNx{9781450380966}
\urldef\tempurl%
\url{https://doi.org/10.1145/3411764.3445695}
\showDOI{\tempurl}


\bibitem[Lafreni{\`{e}}re et~al\mbox{.}(2012)]%
        {Lafrenire2012GAMS}
\bibfield{author}{\bibinfo{person}{Marc-Andr{\'{e}}~K. Lafreni{\`{e}}re},
  \bibinfo{person}{J{\'{e}}r{\'{e}}mie Verner-Filion}, {and}
  \bibinfo{person}{Robert~J. Vallerand}.} \bibinfo{year}{2012}\natexlab{}.
\newblock \showarticletitle{Development and validation of the Gaming Motivation
  Scale ({GAMS})}.
\newblock \bibinfo{journal}{\emph{Personality and Individual Differences}}
  \bibinfo{volume}{53}, \bibinfo{number}{7} (\bibinfo{date}{Nov.}
  \bibinfo{year}{2012}), \bibinfo{pages}{827--831}.
\newblock
\urldef\tempurl%
\url{https://doi.org/10.1016/j.paid.2012.06.013}
\showDOI{\tempurl}


\bibitem[Law et~al\mbox{.}(2018)]%
        {sys2}
\bibfield{author}{\bibinfo{person}{Effie L.-C. Law}, \bibinfo{person}{Florian
  Br\"{u}hlmann}, {and} \bibinfo{person}{Elisa~D. Mekler}.}
  \bibinfo{year}{2018}\natexlab{}.
\newblock \showarticletitle{Systematic Review and Validation of the Game
  Experience Questionnaire (GEQ) - Implications for Citation and Reporting
  Practice}. In \bibinfo{booktitle}{\emph{Proceedings of the 2018 Annual
  Symposium on Computer-Human Interaction in Play}} (<conf-loc>,
  <city>Melbourne</city>, <state>VIC</state>, <country>Australia</country>,
  </conf-loc>) \emph{(\bibinfo{series}{CHI PLAY '18})}.
  \bibinfo{publisher}{Association for Computing Machinery},
  \bibinfo{address}{New York, NY, USA}, \bibinfo{pages}{257–270}.
\newblock
\showISBNx{9781450356244}
\urldef\tempurl%
\url{https://doi.org/10.1145/3242671.3242683}
\showDOI{\tempurl}


\bibitem[Lehtonen et~al\mbox{.}(2019)]%
        {Lehtonen}
\bibfield{author}{\bibinfo{person}{Lauri Lehtonen}, \bibinfo{person}{Maximus~D.
  Kaos}, \bibinfo{person}{Raine Kajastila}, \bibinfo{person}{Leo Holsti},
  \bibinfo{person}{Janne Karsisto}, \bibinfo{person}{Sami Pekkola},
  \bibinfo{person}{Joni V\"{a}h\"{a}m\"{a}ki}, \bibinfo{person}{Lassi
  Vapaakallio}, {and} \bibinfo{person}{Perttu H\"{a}m\"{a}l\"{a}inen}.}
  \bibinfo{year}{2019}\natexlab{}.
\newblock \showarticletitle{Movement Empowerment in a Multiplayer Mixed-Reality
  Trampoline Game}. In \bibinfo{booktitle}{\emph{Proceedings of the Annual
  Symposium on Computer-Human Interaction in Play}} (Barcelona, Spain)
  \emph{(\bibinfo{series}{CHI PLAY '19})}. \bibinfo{publisher}{Association for
  Computing Machinery}, \bibinfo{address}{New York, NY, USA},
  \bibinfo{pages}{19–29}.
\newblock
\showISBNx{9781450366885}
\urldef\tempurl%
\url{https://doi.org/10.1145/3311350.3347181}
\showDOI{\tempurl}


\bibitem[Lerch et~al\mbox{.}(2018)]%
        {lerch}
\bibfield{author}{\bibinfo{person}{Vanessa~R. Lerch},
  \bibinfo{person}{Sharon~T. Steinemann}, {and} \bibinfo{person}{Klaus Opwis}.}
  \bibinfo{year}{2018}\natexlab{}.
\newblock \showarticletitle{Understanding Fitness App Usage Over Time: Moving
  Beyond the Need for Competence}. In \bibinfo{booktitle}{\emph{Extended
  Abstracts of the 2018 CHI Conference on Human Factors in Computing Systems}}
  (<conf-loc>, <city>Montreal QC</city>, <country>Canada</country>,
  </conf-loc>) \emph{(\bibinfo{series}{CHI EA '18})}.
  \bibinfo{publisher}{Association for Computing Machinery},
  \bibinfo{address}{New York, NY, USA}, \bibinfo{pages}{1–6}.
\newblock
\showISBNx{9781450356213}
\urldef\tempurl%
\url{https://doi.org/10.1145/3170427.3188608}
\showDOI{\tempurl}


\bibitem[Lukoff et~al\mbox{.}(2022)]%
        {Lukoff2022-xs}
\bibfield{author}{\bibinfo{person}{Kai Lukoff}, \bibinfo{person}{Ulrik Lyngs},
  {and} \bibinfo{person}{Lize Alberts}.} \bibinfo{year}{2022}\natexlab{}.
\newblock \showarticletitle{Designing to Support Autonomy and Reduce
  Psychological Reactance in Digital {Self-Control} Tools}. In
  \bibinfo{booktitle}{\emph{Position Papers for the Workshop
  ``{Self-Determination} Theory in {HCI}: Shaping a Research Agenda'' at the
  Conference on Human Factors in Computing Systems ({CHI} '22)}}.
  \bibinfo{publisher}{ACM}, \bibinfo{address}{New Orleans, LA, USA},
  \bibinfo{pages}{5}.
\newblock


\bibitem[Lyngs et~al\mbox{.}(2019)]%
        {Lyngs2019SelfControlInCyberspace}
\bibfield{author}{\bibinfo{person}{Ulrik Lyngs}, \bibinfo{person}{Kai Lukoff},
  \bibinfo{person}{Petr Slovak}, \bibinfo{person}{Reuben Binns},
  \bibinfo{person}{Adam Slack}, \bibinfo{person}{Michael Inzlicht},
  \bibinfo{person}{Max Van~Kleek}, {and} \bibinfo{person}{Nigel Shadbolt}.}
  \bibinfo{year}{2019}\natexlab{}.
\newblock \showarticletitle{Self-Control in Cyberspace: Applying Dual Systems
  Theory to a Review of Digital Self-Control Tools}. In
  \bibinfo{booktitle}{\emph{Proceedings of the 2019 CHI Conference on Human
  Factors in Computing Systems}} (Glasgow, Scotland Uk)
  \emph{(\bibinfo{series}{CHI '19})}. \bibinfo{publisher}{Association for
  Computing Machinery}, \bibinfo{address}{New York, NY, USA},
  \bibinfo{pages}{1–18}.
\newblock
\showISBNx{9781450359702}
\urldef\tempurl%
\url{https://doi.org/10.1145/3290605.3300361}
\showDOI{\tempurl}


\bibitem[Lyngs et~al\mbox{.}(2020)]%
        {Lyngs2020FacebookHack}
\bibfield{author}{\bibinfo{person}{Ulrik Lyngs}, \bibinfo{person}{Kai Lukoff},
  \bibinfo{person}{Petr Slovak}, \bibinfo{person}{William Seymour},
  \bibinfo{person}{Helena Webb}, \bibinfo{person}{Marina Jirotka},
  \bibinfo{person}{Jun Zhao}, \bibinfo{person}{Max Van~Kleek}, {and}
  \bibinfo{person}{Nigel Shadbolt}.} \bibinfo{year}{2020}\natexlab{}.
\newblock \showarticletitle{'I Just Want to Hack Myself to Not Get Distracted':
  Evaluating Design Interventions for {Self-Control} on Facebook}.
\newblock In \bibinfo{booktitle}{\emph{Proceedings of the 2020 {CHI} Conference
  on Human Factors in Computing Systems}}. \bibinfo{publisher}{Association for
  Computing Machinery}, \bibinfo{address}{New York, NY, USA},
  \bibinfo{pages}{1--15}.
\newblock
\showISBNx{9781450367080}
\urldef\tempurl%
\url{https://doi.org/10.1145/3313831.3376672}
\showDOI{\tempurl}


\bibitem[Marques et~al\mbox{.}(2021)]%
        {60}
\bibfield{author}{\bibinfo{person}{Bertil~P. Marques}, \bibinfo{person}{Rosa
  Reis}, {and} \bibinfo{person}{Mar\'{\i}lio Cardoso}.}
  \bibinfo{year}{2021}\natexlab{}.
\newblock \showarticletitle{Games: The Motivation in Engineering Education}. In
  \bibinfo{booktitle}{\emph{Ninth International Conference on Technological
  Ecosystems for Enhancing Multiculturality (TEEM'21)}} (Barcelona, Spain)
  \emph{(\bibinfo{series}{TEEM'21})}. \bibinfo{publisher}{Association for
  Computing Machinery}, \bibinfo{address}{New York, NY, USA},
  \bibinfo{pages}{395–399}.
\newblock
\showISBNx{9781450390668}
\urldef\tempurl%
\url{https://doi.org/10.1145/3486011.3486483}
\showDOI{\tempurl}


\bibitem[McAlaney et~al\mbox{.}(2020)]%
        {McAlaney2020-dc}
\bibfield{author}{\bibinfo{person}{John McAlaney}, \bibinfo{person}{Manal
  Aldhayan}, \bibinfo{person}{Mohamed~Basel Almourad},
  \bibinfo{person}{Sainabou Cham}, {and} \bibinfo{person}{Raian Ali}.}
  \bibinfo{year}{2020}\natexlab{}.
\newblock \showarticletitle{On the Need for Cultural Sensitivity in Digital
  Wellbeing Tools and Messages: A {UK-China} Comparison}. In
  \bibinfo{booktitle}{\emph{Trends and Innovations in Information Systems and
  Technologies}}. \bibinfo{publisher}{Springer International Publishing},
  \bibinfo{address}{Cham}, \bibinfo{pages}{723--733}.
\newblock
\urldef\tempurl%
\url{https://doi.org/10.1007/978-3-030-45691-7\_68}
\showDOI{\tempurl}


\bibitem[Mekler et~al\mbox{.}(2014)]%
        {sys3}
\bibfield{author}{\bibinfo{person}{Elisa~D. Mekler},
  \bibinfo{person}{Julia~Ayumi Bopp}, \bibinfo{person}{Alexandre~N. Tuch},
  {and} \bibinfo{person}{Klaus Opwis}.} \bibinfo{year}{2014}\natexlab{}.
\newblock \showarticletitle{A systematic review of quantitative studies on the
  enjoyment of digital entertainment games}. In
  \bibinfo{booktitle}{\emph{Proceedings of the SIGCHI Conference on Human
  Factors in Computing Systems}} (Toronto, Ontario, Canada)
  \emph{(\bibinfo{series}{CHI '14})}. \bibinfo{publisher}{Association for
  Computing Machinery}, \bibinfo{address}{New York, NY, USA},
  \bibinfo{pages}{927–936}.
\newblock
\showISBNx{9781450324731}
\urldef\tempurl%
\url{https://doi.org/10.1145/2556288.2557078}
\showDOI{\tempurl}


\bibitem[Michie and Johnston(2012)]%
        {michie}
\bibfield{author}{\bibinfo{person}{Susan Michie} {and} \bibinfo{person}{Marie
  Johnston}.} \bibinfo{year}{2012}\natexlab{}.
\newblock \showarticletitle{Theories and techniques of behaviour change:
  Developing a cumulative science of behaviour change}.
\newblock \bibinfo{journal}{\emph{Health Psychology Review}}
  \bibinfo{volume}{6}, \bibinfo{number}{1} (\bibinfo{year}{2012}),
  \bibinfo{pages}{1--6}.
\newblock
\urldef\tempurl%
\url{https://doi.org/10.1080/17437199.2012.654964}
\showDOI{\tempurl}
\showeprint{https://doi.org/10.1080/17437199.2012.654964}


\bibitem[Michie and Prestwich(2010)]%
        {michie2010interventions}
\bibfield{author}{\bibinfo{person}{Susan Michie} {and} \bibinfo{person}{Andrew
  Prestwich}.} \bibinfo{year}{2010}\natexlab{}.
\newblock \showarticletitle{Are interventions theory-based? Development of a
  theory coding scheme.}
\newblock \bibinfo{journal}{\emph{Health psychology}} \bibinfo{volume}{29},
  \bibinfo{number}{1} (\bibinfo{year}{2010}), \bibinfo{pages}{1}.
\newblock


\bibitem[Mikulic(2022)]%
        {statista}
\bibfield{author}{\bibinfo{person}{M Mikulic}.}
  \bibinfo{year}{2022}\natexlab{}.
\newblock \bibinfo{booktitle}{\emph{Number of mHealth apps available in the
  Apple App Store from 1st quarter 2015 to 3rd quarter 2022}}.
\newblock \bibinfo{type}{{T}echnical {R}eport}.
  \bibinfo{institution}{Statista}.
\newblock
\urldef\tempurl%
\url{https://www.statista.com/topics/2263/mhealth}
\showURL{%
\tempurl}


\bibitem[Molina et~al\mbox{.}(2023)]%
        {Molina}
\bibfield{author}{\bibinfo{person}{Maria~D. Molina}, \bibinfo{person}{Emily~S
  Zhan}, \bibinfo{person}{Devanshi Agnihotri}, \bibinfo{person}{Saeed
  Abdullah}, {and} \bibinfo{person}{Pallav Deka}.}
  \bibinfo{year}{2023}\natexlab{}.
\newblock \showarticletitle{Motivation to Use Fitness Application for Improving
  Physical Activity Among Hispanic Users: The Pivotal Role of Interactivity and
  Relatedness}. In \bibinfo{booktitle}{\emph{Proceedings of the 2023 CHI
  Conference on Human Factors in Computing Systems}} (Hamburg, Germany)
  \emph{(\bibinfo{series}{CHI '23})}. \bibinfo{publisher}{Association for
  Computing Machinery}, \bibinfo{address}{New York, NY, USA}, Article
  \bibinfo{articleno}{321}, \bibinfo{numpages}{13}~pages.
\newblock
\showISBNx{9781450394215}
\urldef\tempurl%
\url{https://doi.org/10.1145/3544548.3581200}
\showDOI{\tempurl}


\bibitem[Mustafa et~al\mbox{.}(2022)]%
        {mustafa2022user}
\bibfield{author}{\bibinfo{person}{Abdulsalam~Salihu Mustafa},
  \bibinfo{person}{Nor’ashikin Ali}, \bibinfo{person}{Jaspaljeet~Singh
  Dhillon}, \bibinfo{person}{Gamal Alkawsi}, {and} \bibinfo{person}{Yahia
  Baashar}.} \bibinfo{year}{2022}\natexlab{}.
\newblock \showarticletitle{User engagement and abandonment of mHealth: a
  cross-sectional survey}.
\newblock \bibinfo{journal}{\emph{Healthcare}} \bibinfo{volume}{10},
  \bibinfo{number}{2} (\bibinfo{year}{2022}), \bibinfo{pages}{221}.
\newblock


\bibitem[Naderi et~al\mbox{.}(2014)]%
        {54}
\bibfield{author}{\bibinfo{person}{Babak Naderi}, \bibinfo{person}{Ina
  Wechsung}, \bibinfo{person}{Tim Polzehl}, {and} \bibinfo{person}{Sebastian
  M\"{o}ller}.} \bibinfo{year}{2014}\natexlab{}.
\newblock \showarticletitle{Development and Validation of Extrinsic Motivation
  Scale for Crowdsourcing Micro-Task Platforms}. In
  \bibinfo{booktitle}{\emph{Proceedings of the 2014 International ACM Workshop
  on Crowdsourcing for Multimedia}} (Orlando, Florida, USA)
  \emph{(\bibinfo{series}{CrowdMM '14})}. \bibinfo{publisher}{Association for
  Computing Machinery}, \bibinfo{address}{New York, NY, USA},
  \bibinfo{pages}{31–36}.
\newblock
\showISBNx{9781450331289}
\urldef\tempurl%
\url{https://doi.org/10.1145/2660114.2660122}
\showDOI{\tempurl}


\bibitem[Ntoumanis et~al\mbox{.}(2021)]%
        {metantoumanis}
\bibfield{author}{\bibinfo{person}{Nikos Ntoumanis}, \bibinfo{person}{Johan~YY
  Ng}, \bibinfo{person}{Andrew Prestwich}, \bibinfo{person}{Eleanor Quested},
  \bibinfo{person}{Jennie~E Hancox}, \bibinfo{person}{Cecilie
  Th{\o}gersen-Ntoumani}, \bibinfo{person}{Edward~L Deci},
  \bibinfo{person}{Richard~M Ryan}, \bibinfo{person}{Chris Lonsdale}, {and}
  \bibinfo{person}{Geoffrey~C Williams}.} \bibinfo{year}{2021}\natexlab{}.
\newblock \showarticletitle{A meta-analysis of self-determination
  theory-informed intervention studies in the health domain: Effects on
  motivation, health behavior, physical, and psychological health}.
\newblock \bibinfo{journal}{\emph{Health psychology review}}
  \bibinfo{volume}{15}, \bibinfo{number}{2} (\bibinfo{year}{2021}),
  \bibinfo{pages}{214--244}.
\newblock


\bibitem[Pelletier and Sharp(2008)]%
        {pelletier2008persuasive}
\bibfield{author}{\bibinfo{person}{Luc~G Pelletier} {and}
  \bibinfo{person}{Elizabeth Sharp}.} \bibinfo{year}{2008}\natexlab{}.
\newblock \showarticletitle{Persuasive communication and proenvironmental
  behaviours: how message tailoring and message framing can improve the
  integration of behaviours through self-determined motivation.}
\newblock \bibinfo{journal}{\emph{Canadian Psychology/Psychologie Canadienne}}
  \bibinfo{volume}{49}, \bibinfo{number}{3} (\bibinfo{year}{2008}),
  \bibinfo{pages}{210}.
\newblock


\bibitem[Peters et~al\mbox{.}(2018)]%
        {Peters2018-ra}
\bibfield{author}{\bibinfo{person}{Dorian Peters}, \bibinfo{person}{Rafael~A
  Calvo}, {and} \bibinfo{person}{Richard~M Ryan}.}
  \bibinfo{year}{2018}\natexlab{}.
\newblock \showarticletitle{Designing for Motivation, Engagement and Wellbeing
  in Digital Experience}.
\newblock \bibinfo{journal}{\emph{Frontiers in psychology}}
  \bibinfo{volume}{9} (\bibinfo{date}{May} \bibinfo{year}{2018}),
  \bibinfo{pages}{797}.
\newblock
\showISSN{1664-1078}
\urldef\tempurl%
\url{https://doi.org/10.3389/fpsyg.2018.00797}
\showDOI{\tempurl}


\bibitem[Pinder et~al\mbox{.}(2018)]%
        {Pinder}
\bibfield{author}{\bibinfo{person}{Charlie Pinder}, \bibinfo{person}{Jo
  Vermeulen}, \bibinfo{person}{Benjamin~R. Cowan}, {and}
  \bibinfo{person}{Russell Beale}.} \bibinfo{year}{2018}\natexlab{}.
\newblock \showarticletitle{Digital Behaviour Change Interventions to Break and
  Form Habits}.
\newblock \bibinfo{journal}{\emph{ACM Trans. Comput.-Hum. Interact.}}
  \bibinfo{volume}{25}, \bibinfo{number}{3}, Article \bibinfo{articleno}{15}
  (\bibinfo{date}{jun} \bibinfo{year}{2018}), \bibinfo{numpages}{66}~pages.
\newblock
\showISSN{1073-0516}
\urldef\tempurl%
\url{https://doi.org/10.1145/3196830}
\showDOI{\tempurl}


\bibitem[Poeller and Phillips(2022)]%
        {34}
\bibfield{author}{\bibinfo{person}{Susanne Poeller} {and}
  \bibinfo{person}{Cody~J. Phillips}.} \bibinfo{year}{2022}\natexlab{}.
\newblock \showarticletitle{Self-Determination Theory — I Choose You! The
  Limitations of Viewing Motivation in HCI Research Through the Lens of a
  Single Theory}. In \bibinfo{booktitle}{\emph{Extended Abstracts of the 2022
  Annual Symposium on Computer-Human Interaction in Play}} (Bremen, Germany)
  \emph{(\bibinfo{series}{CHI PLAY '22})}. \bibinfo{publisher}{Association for
  Computing Machinery}, \bibinfo{address}{New York, NY, USA},
  \bibinfo{pages}{261–262}.
\newblock
\showISBNx{9781450392112}
\urldef\tempurl%
\url{https://doi.org/10.1145/3505270.3558361}
\showDOI{\tempurl}


\bibitem[Posch et~al\mbox{.}(2019)]%
        {20}
\bibfield{author}{\bibinfo{person}{Lisa Posch}, \bibinfo{person}{Arnim Bleier},
  \bibinfo{person}{Clemens~M. Lechner}, \bibinfo{person}{Daniel Danner},
  \bibinfo{person}{Fabian Fl\"{o}ck}, {and} \bibinfo{person}{Markus
  Strohmaier}.} \bibinfo{year}{2019}\natexlab{}.
\newblock \showarticletitle{Measuring Motivations of Crowdworkers: The
  Multidimensional Crowdworker Motivation Scale}.
\newblock \bibinfo{journal}{\emph{Trans. Soc. Comput.}} \bibinfo{volume}{2},
  \bibinfo{number}{2}, Article \bibinfo{articleno}{8} (\bibinfo{date}{sep}
  \bibinfo{year}{2019}), \bibinfo{numpages}{34}~pages.
\newblock
\showISSN{2469-7818}
\urldef\tempurl%
\url{https://doi.org/10.1145/3335081}
\showDOI{\tempurl}


\bibitem[Prochaska and Velicer(1997)]%
        {trans}
\bibfield{author}{\bibinfo{person}{James~O Prochaska} {and}
  \bibinfo{person}{Wayne~F Velicer}.} \bibinfo{year}{1997}\natexlab{}.
\newblock \showarticletitle{The transtheoretical model of health behavior
  change}.
\newblock \bibinfo{journal}{\emph{American journal of health promotion}}
  \bibinfo{volume}{12}, \bibinfo{number}{1} (\bibinfo{year}{1997}),
  \bibinfo{pages}{38--48}.
\newblock


\bibitem[Putnam et~al\mbox{.}(2017)]%
        {Putnam}
\bibfield{author}{\bibinfo{person}{Cynthia Putnam}, \bibinfo{person}{Amanda
  Lin}, \bibinfo{person}{Vansanth Subramanian}, \bibinfo{person}{Dorian~C.
  Anderson}, \bibinfo{person}{Erica Christian}, \bibinfo{person}{Bharathi
  Swaminathan}, \bibinfo{person}{Sai Yalla}, \bibinfo{person}{William Cotter},
  \bibinfo{person}{Danielle Ciccone}, {and} \bibinfo{person}{Jinghui Cheng}.}
  \bibinfo{year}{2017}\natexlab{}.
\newblock \showarticletitle{Effects of Commercial Exergames on Motivation in
  Brian Injury Therapy}. In \bibinfo{booktitle}{\emph{Extended Abstracts
  Publication of the Annual Symposium on Computer-Human Interaction in Play}}
  (Amsterdam, The Netherlands) \emph{(\bibinfo{series}{CHI PLAY '17 Extended
  Abstracts})}. \bibinfo{publisher}{Association for Computing Machinery},
  \bibinfo{address}{New York, NY, USA}, \bibinfo{pages}{47–59}.
\newblock
\showISBNx{9781450351119}
\urldef\tempurl%
\url{https://doi.org/10.1145/3130859.3131431}
\showDOI{\tempurl}


\bibitem[Roffarello and De~Russis(2023)]%
        {roff}
\bibfield{author}{\bibinfo{person}{Alberto~Monge Roffarello} {and}
  \bibinfo{person}{Luigi De~Russis}.} \bibinfo{year}{2023}\natexlab{}.
\newblock \showarticletitle{Achieving Digital Wellbeing Through Digital
  Self-Control Tools: A Systematic Review and Meta-Analysis}.
\newblock \bibinfo{journal}{\emph{ACM Trans. Comput.-Hum. Interact.}}
  \bibinfo{volume}{30}, \bibinfo{number}{4}, Article \bibinfo{articleno}{53}
  (\bibinfo{date}{sep} \bibinfo{year}{2023}), \bibinfo{numpages}{66}~pages.
\newblock
\showISSN{1073-0516}
\urldef\tempurl%
\url{https://doi.org/10.1145/3571810}
\showDOI{\tempurl}


\bibitem[Ryan and Deci(2000a)]%
        {ryan2000intrinsic}
\bibfield{author}{\bibinfo{person}{Richard~M Ryan} {and}
  \bibinfo{person}{Edward~L Deci}.} \bibinfo{year}{2000}\natexlab{a}.
\newblock \showarticletitle{Intrinsic and extrinsic motivations: Classic
  definitions and new directions}.
\newblock \bibinfo{journal}{\emph{Contemporary educational psychology}}
  \bibinfo{volume}{25}, \bibinfo{number}{1} (\bibinfo{year}{2000}),
  \bibinfo{pages}{54--67}.
\newblock


\bibitem[Ryan and Deci(2000b)]%
        {ryan2000self}
\bibfield{author}{\bibinfo{person}{Richard~M Ryan} {and}
  \bibinfo{person}{Edward~L Deci}.} \bibinfo{year}{2000}\natexlab{b}.
\newblock \showarticletitle{Self-determination theory and the facilitation of
  intrinsic motivation, social development, and well-being.}
\newblock \bibinfo{journal}{\emph{American psychologist}} \bibinfo{volume}{55},
  \bibinfo{number}{1} (\bibinfo{year}{2000}), \bibinfo{pages}{68}.
\newblock


\bibitem[Ryan and Deci(2012)]%
        {ryan2012multiple}
\bibfield{author}{\bibinfo{person}{Richard~M Ryan} {and}
  \bibinfo{person}{Edward~L Deci}.} \bibinfo{year}{2012}\natexlab{}.
\newblock \showarticletitle{Multiple identities within a single self}.
\newblock In \bibinfo{booktitle}{\emph{Handbook of self and identity}}.
  \bibinfo{publisher}{Guilford Press}, \bibinfo{address}{New York, NY},
  \bibinfo{pages}{225--246}.
\newblock


\bibitem[Ryan and Deci(2016)]%
        {ryan2016facilitating}
\bibfield{author}{\bibinfo{person}{Richard~M Ryan} {and}
  \bibinfo{person}{Edward~L Deci}.} \bibinfo{year}{2016}\natexlab{}.
\newblock \showarticletitle{Facilitating and hindering motivation, learning,
  and well-being in schools}.
\newblock In \bibinfo{booktitle}{\emph{Handbook of motivation at school}}.
  \bibinfo{publisher}{Routledge}, \bibinfo{address}{New York, NY},
  \bibinfo{pages}{96--119}.
\newblock


\bibitem[Ryan and Deci(2017)]%
        {ryan2017self}
\bibfield{author}{\bibinfo{person}{Richard~M Ryan} {and}
  \bibinfo{person}{Edward~L Deci}.} \bibinfo{year}{2017}\natexlab{}.
\newblock \bibinfo{booktitle}{\emph{Self-determination theory: Basic
  psychological needs in motivation, development, and wellness}}.
\newblock \bibinfo{publisher}{The Guilford Press}, \bibinfo{address}{New York,
  NY}.
\newblock
\urldef\tempurl%
\url{https://doi.org/10.1521/978.14625/28806}
\showURL{%
\tempurl}


\bibitem[Ryan et~al\mbox{.}(2022)]%
        {ryan2022we}
\bibfield{author}{\bibinfo{person}{Richard~M Ryan}, \bibinfo{person}{Jasper~J
  Duineveld}, \bibinfo{person}{Stefano~I Di~Domenico},
  \bibinfo{person}{William~S Ryan}, \bibinfo{person}{Ben~A Steward}, {and}
  \bibinfo{person}{Emma~L Bradshaw}.} \bibinfo{year}{2022}\natexlab{}.
\newblock \showarticletitle{We know this much is (meta-analytically) true: A
  meta-review of meta-analytic findings evaluating self-determination theory.}
\newblock \bibinfo{journal}{\emph{Psychological Bulletin}}
  \bibinfo{volume}{148}, \bibinfo{number}{11-12} (\bibinfo{year}{2022}),
  \bibinfo{pages}{813}.
\newblock


\bibitem[Ryan et~al\mbox{.}(1983)]%
        {imi}
\bibfield{author}{\bibinfo{person}{Richard~M Ryan}, \bibinfo{person}{Valerie
  Mims}, {and} \bibinfo{person}{Richard Koestner}.}
  \bibinfo{year}{1983}\natexlab{}.
\newblock \showarticletitle{Relation of reward contingency and interpersonal
  context to intrinsic motivation: A review and test using cognitive evaluation
  theory.}
\newblock \bibinfo{journal}{\emph{Journal of personality and Social
  Psychology}} \bibinfo{volume}{45}, \bibinfo{number}{4}
  (\bibinfo{year}{1983}), \bibinfo{pages}{736}.
\newblock


\bibitem[Saksono et~al\mbox{.}(2020)]%
        {Saksono}
\bibfield{author}{\bibinfo{person}{Herman Saksono}, \bibinfo{person}{Carmen
  Castaneda-Sceppa}, \bibinfo{person}{Jessica Hoffman}, \bibinfo{person}{Vivien
  Morris}, \bibinfo{person}{Magy Seif El-Nasr}, {and}
  \bibinfo{person}{Andrea~G. Parker}.} \bibinfo{year}{2020}\natexlab{}.
\newblock \showarticletitle{Storywell: Designing for Family Fitness App
  Motivation by Using Social Rewards and Reflection}. In
  \bibinfo{booktitle}{\emph{Proceedings of the 2020 CHI Conference on Human
  Factors in Computing Systems}} (Honolulu, HI, USA)
  \emph{(\bibinfo{series}{CHI '20})}. \bibinfo{publisher}{Association for
  Computing Machinery}, \bibinfo{address}{New York, NY, USA},
  \bibinfo{pages}{1–13}.
\newblock
\showISBNx{9781450367080}
\urldef\tempurl%
\url{https://doi.org/10.1145/3313831.3376686}
\showDOI{\tempurl}


\bibitem[Schmidt-Kraepelin et~al\mbox{.}(2019)]%
        {schmidt2019gamification}
\bibfield{author}{\bibinfo{person}{Manuel Schmidt-Kraepelin},
  \bibinfo{person}{Scott Thiebes}, \bibinfo{person}{Stefan Stepanovic},
  \bibinfo{person}{Tobias Mettler}, {and} \bibinfo{person}{Ali Sunyaev}.}
  \bibinfo{year}{2019}\natexlab{}.
\newblock \showarticletitle{Gamification in Health Behavior Change Support
  Systems - A Synthesis of Unintended Side Effects}. In
  \bibinfo{booktitle}{\emph{Wirtschaftsinformatik}}. \bibinfo{publisher}{14th
  International Conference on Wirtschaftsinformatik}, \bibinfo{address}{Siegen,
  Germany}, \bibinfo{pages}{1032--1046}.
\newblock
\urldef\tempurl%
\url{https://api.semanticscholar.org/CorpusID:145042569}
\showURL{%
\tempurl}


\bibitem[Sheldon and Hilpert(2012)]%
        {sheldon2012balanced}
\bibfield{author}{\bibinfo{person}{Kennon~M Sheldon} {and}
  \bibinfo{person}{Jonathan~C Hilpert}.} \bibinfo{year}{2012}\natexlab{}.
\newblock \showarticletitle{The balanced measure of psychological needs (BMPN)
  scale: An alternative domain general measure of need satisfaction}.
\newblock \bibinfo{journal}{\emph{Motivation and Emotion}}
  \bibinfo{volume}{36} (\bibinfo{year}{2012}), \bibinfo{pages}{439--451}.
\newblock


\bibitem[Sinai and Rosenberg-Kima(2022)]%
        {Sinai}
\bibfield{author}{\bibinfo{person}{Dafna Sinai} {and} \bibinfo{person}{Rinat~B.
  Rosenberg-Kima}.} \bibinfo{year}{2022}\natexlab{}.
\newblock \showarticletitle{Perceptions of Social Robots as Motivating Learning
  Companions for Online Learning}. In \bibinfo{booktitle}{\emph{Proceedings of
  the 2022 ACM/IEEE International Conference on Human-Robot Interaction}}
  \emph{(\bibinfo{series}{HRI '22})}. \bibinfo{publisher}{IEEE Press},
  \bibinfo{address}{Sapporo, Hokkaido, Japan}, \bibinfo{pages}{1045–1048}.
\newblock


\bibitem[Spiel et~al\mbox{.}(2018)]%
        {spiel2018fitter}
\bibfield{author}{\bibinfo{person}{Katta Spiel}, \bibinfo{person}{Fares
  Kayali}, \bibinfo{person}{Louise Horvath}, \bibinfo{person}{Michael Penkler},
  \bibinfo{person}{Sabine Harrer}, \bibinfo{person}{Miguel Sicart}, {and}
  \bibinfo{person}{Jessica Hammer}.} \bibinfo{year}{2018}\natexlab{}.
\newblock \showarticletitle{Fitter, Happier, More Productive? The Normative
  Ontology of Fitness Trackers}. In \bibinfo{booktitle}{\emph{Extended
  Abstracts of the 2018 CHI Conference on Human Factors in Computing Systems}}
  (<conf-loc>, <city>Montreal QC</city>, <country>Canada</country>,
  </conf-loc>) \emph{(\bibinfo{series}{CHI EA '18})}.
  \bibinfo{publisher}{Association for Computing Machinery},
  \bibinfo{address}{New York, NY, USA}, \bibinfo{pages}{1–10}.
\newblock
\showISBNx{9781450356213}
\urldef\tempurl%
\url{https://doi.org/10.1145/3170427.3188401}
\showDOI{\tempurl}


\bibitem[Stibe and Cugelman(2016)]%
        {stibe2016persuasive}
\bibfield{author}{\bibinfo{person}{Agnis Stibe} {and} \bibinfo{person}{Brian
  Cugelman}.} \bibinfo{year}{2016}\natexlab{}.
\newblock \showarticletitle{Persuasive backfiring: When behavior change
  interventions trigger unintended negative outcomes}. In
  \bibinfo{booktitle}{\emph{Persuasive Technology: Proceedings of the 11th
  International Conference}}. Springer, \bibinfo{publisher}{PERSUASIVE 2016},
  \bibinfo{address}{Salzburg, Austria}, \bibinfo{pages}{65--77}.
\newblock


\bibitem[Sundar et~al\mbox{.}(2012)]%
        {sundar2012motivational}
\bibfield{author}{\bibinfo{person}{S.~Shyam Sundar},
  \bibinfo{person}{Saraswathi Bellur}, {and} \bibinfo{person}{Haiyan Jia}.}
  \bibinfo{year}{2012}\natexlab{}.
\newblock \showarticletitle{Motivational technologies: a theoretical framework
  for designing preventive health applications}. In
  \bibinfo{booktitle}{\emph{Design for Health and Safety: Proceedings of the
  7th International Conference}}. Springer, \bibinfo{publisher}{PERSUASIVE
  2012}, \bibinfo{address}{Link{\"o}ping, Sweden}, \bibinfo{pages}{112--122}.
\newblock


\bibitem[Tolentino and Roleda(2019)]%
        {16}
\bibfield{author}{\bibinfo{person}{Analyn~N. Tolentino} {and}
  \bibinfo{person}{Lydia~S. Roleda}.} \bibinfo{year}{2019}\natexlab{}.
\newblock \showarticletitle{Gamified Physics Instruction in a Reformatory
  Classroom Context}. In \bibinfo{booktitle}{\emph{Proceedings of the 10th
  International Conference on E-Education, E-Business, E-Management and
  E-Learning}} (Tokyo, Japan) \emph{(\bibinfo{series}{IC4E '19})}.
  \bibinfo{publisher}{Association for Computing Machinery},
  \bibinfo{address}{New York, NY, USA}, \bibinfo{pages}{135–140}.
\newblock
\showISBNx{9781450366021}
\urldef\tempurl%
\url{https://doi.org/10.1145/3306500.3306527}
\showDOI{\tempurl}


\bibitem[Tsay et~al\mbox{.}(2020)]%
        {tsay2020overcoming}
\bibfield{author}{\bibinfo{person}{Crystal Han-Huei Tsay},
  \bibinfo{person}{Alexander~K Kofinas}, \bibinfo{person}{Smita~K Trivedi},
  {and} \bibinfo{person}{Yang Yang}.} \bibinfo{year}{2020}\natexlab{}.
\newblock \showarticletitle{Overcoming the novelty effect in online gamified
  learning systems: An empirical evaluation of student engagement and
  performance}.
\newblock \bibinfo{journal}{\emph{Journal of Computer Assisted Learning}}
  \bibinfo{volume}{36}, \bibinfo{number}{2} (\bibinfo{year}{2020}),
  \bibinfo{pages}{128--146}.
\newblock


\bibitem[Tyack and Mekler(2020)]%
        {games}
\bibfield{author}{\bibinfo{person}{April Tyack} {and} \bibinfo{person}{Elisa~D.
  Mekler}.} \bibinfo{year}{2020}\natexlab{}.
\newblock \bibinfo{booktitle}{\emph{Self-Determination Theory in HCI Games
  Research: Current Uses and Open Questions}}.
\newblock \bibinfo{publisher}{Association for Computing Machinery},
  \bibinfo{address}{New York, NY, USA}, \bibinfo{pages}{1–22}.
\newblock
\showISBNx{9781450367080}
\urldef\tempurl%
\url{https://doi.org/10.1145/3313831.3376723}
\showURL{%
\tempurl}


\bibitem[Vallerand(1997)]%
        {vallerand1997toward}
\bibfield{author}{\bibinfo{person}{Robert~J Vallerand}.}
  \bibinfo{year}{1997}\natexlab{}.
\newblock \showarticletitle{Toward a hierarchical model of intrinsic and
  extrinsic motivation}.
\newblock In \bibinfo{booktitle}{\emph{Advances in experimental social
  psychology}}. Vol.~\bibinfo{volume}{29}. \bibinfo{publisher}{Elsevier},
  \bibinfo{address}{Amsterdam, The Netherlands}, \bibinfo{pages}{271--360}.
\newblock


\bibitem[van Minkelen et~al\mbox{.}(2020)]%
        {Minkelen}
\bibfield{author}{\bibinfo{person}{Peggy van Minkelen}, \bibinfo{person}{Carmen
  Gruson}, \bibinfo{person}{Pleun van Hees}, \bibinfo{person}{Mirle Willems},
  \bibinfo{person}{Jan de Wit}, \bibinfo{person}{Rian Aarts},
  \bibinfo{person}{Jaap Denissen}, {and} \bibinfo{person}{Paul Vogt}.}
  \bibinfo{year}{2020}\natexlab{}.
\newblock \showarticletitle{Using Self-Determination Theory in Social Robots to
  Increase Motivation in L2 Word Learning}. In
  \bibinfo{booktitle}{\emph{Proceedings of the 2020 ACM/IEEE International
  Conference on Human-Robot Interaction}} (Cambridge, United Kingdom)
  \emph{(\bibinfo{series}{HRI '20})}. \bibinfo{publisher}{Association for
  Computing Machinery}, \bibinfo{address}{New York, NY, USA},
  \bibinfo{pages}{369–377}.
\newblock
\showISBNx{9781450367462}
\urldef\tempurl%
\url{https://doi.org/10.1145/3319502.3374828}
\showDOI{\tempurl}


\bibitem[Villalobos-Z\'{u}\~{n}iga et~al\mbox{.}(2021)]%
        {Villalobos}
\bibfield{author}{\bibinfo{person}{Gabriela Villalobos-Z\'{u}\~{n}iga},
  \bibinfo{person}{Iyubanit Rodr\'{\i}guez}, \bibinfo{person}{Anton Fedosov},
  {and} \bibinfo{person}{Mauro Cherubini}.} \bibinfo{year}{2021}\natexlab{}.
\newblock \showarticletitle{Informed Choices, Progress Monitoring and
  Comparison with Peers: Features to Support the Autonomy, Competence and
  Relatedness Needs, as Suggested by the Self-Determination Theory}. In
  \bibinfo{booktitle}{\emph{Proceedings of the 23rd International Conference on
  Mobile Human-Computer Interaction}} (Toulouse and Virtual, France)
  \emph{(\bibinfo{series}{MobileHCI '21})}. \bibinfo{publisher}{Association for
  Computing Machinery}, \bibinfo{address}{New York, NY, USA}, Article
  \bibinfo{articleno}{13}, \bibinfo{numpages}{14}~pages.
\newblock
\showISBNx{9781450383288}
\urldef\tempurl%
\url{https://doi.org/10.1145/3447526.3472039}
\showDOI{\tempurl}


\bibitem[Villalobos-Z{\'u}{\~n}iga and Cherubini(2020)]%
        {villa}
\bibfield{author}{\bibinfo{person}{Gabriela Villalobos-Z{\'u}{\~n}iga} {and}
  \bibinfo{person}{Mauro Cherubini}.} \bibinfo{year}{2020}\natexlab{}.
\newblock \showarticletitle{Apps that motivate: A taxonomy of app features
  based on self-determination theory}.
\newblock \bibinfo{journal}{\emph{International Journal of Human-Computer
  Studies}}  \bibinfo{volume}{140} (\bibinfo{year}{2020}),
  \bibinfo{pages}{102449}.
\newblock


\bibitem[Williams et~al\mbox{.}(2004)]%
        {williams2004testing}
\bibfield{author}{\bibinfo{person}{Geoffrey~C Williams},
  \bibinfo{person}{Holly~A McGregor}, \bibinfo{person}{Allan Zeldman},
  \bibinfo{person}{Zachary~R Freedman}, {and} \bibinfo{person}{Edward~L Deci}.}
  \bibinfo{year}{2004}\natexlab{}.
\newblock \showarticletitle{Testing a self-determination theory process model
  for promoting glycemic control through diabetes self-management.}
\newblock \bibinfo{journal}{\emph{Health Psychology}} \bibinfo{volume}{23},
  \bibinfo{number}{1} (\bibinfo{year}{2004}), \bibinfo{pages}{58}.
\newblock


\bibitem[Williams et~al\mbox{.}(2009)]%
        {williams2009importance}
\bibfield{author}{\bibinfo{person}{Geoffrey~C Williams},
  \bibinfo{person}{Christopher~P Niemiec}, \bibinfo{person}{Heather Patrick},
  \bibinfo{person}{Richard~M Ryan}, {and} \bibinfo{person}{Edward~L Deci}.}
  \bibinfo{year}{2009}\natexlab{}.
\newblock \showarticletitle{The importance of supporting autonomy and perceived
  competence in facilitating long-term tobacco abstinence}.
\newblock \bibinfo{journal}{\emph{Annals of Behavioral Medicine}}
  \bibinfo{volume}{37}, \bibinfo{number}{3} (\bibinfo{year}{2009}),
  \bibinfo{pages}{315--324}.
\newblock


\bibitem[Wilson et~al\mbox{.}(2006)]%
        {wilson2006formative}
\bibfield{author}{\bibinfo{person}{Dawn~K Wilson}, \bibinfo{person}{Sarah
  Griffin}, \bibinfo{person}{Ruth~P Saunders}, \bibinfo{person}{Alexandra
  Evans}, \bibinfo{person}{Gary Mixon}, \bibinfo{person}{Marcie Wright},
  \bibinfo{person}{Amelia Beasley}, \bibinfo{person}{M~Renee Umstattd},
  \bibinfo{person}{Diana Lattimore}, \bibinfo{person}{Ashley Watts},
  {et~al\mbox{.}}} \bibinfo{year}{2006}\natexlab{}.
\newblock \showarticletitle{Formative evaluation of a motivational intervention
  for increasing physical activity in underserved youth}.
\newblock \bibinfo{journal}{\emph{Evaluation and Program Planning}}
  \bibinfo{volume}{29}, \bibinfo{number}{3} (\bibinfo{year}{2006}),
  \bibinfo{pages}{260--268}.
\newblock


\bibitem[Yang et~al\mbox{.}(2022)]%
        {Yangg}
\bibfield{author}{\bibinfo{person}{Migyeong Yang}, \bibinfo{person}{Kyungha
  Lee}, \bibinfo{person}{Eunji Kim}, \bibinfo{person}{Yeosol Song},
  \bibinfo{person}{Sewang Lee}, \bibinfo{person}{Jiwon Kang},
  \bibinfo{person}{Jinyoung Han}, \bibinfo{person}{Hayeon Song}, {and}
  \bibinfo{person}{Taeeun Kim}.} \bibinfo{year}{2022}\natexlab{}.
\newblock \showarticletitle{Magic Brush: An AI-Based Service for Dementia
  Prevention Focused on Intrinsic Motivation}.
\newblock \bibinfo{journal}{\emph{Proc. ACM Hum.-Comput. Interact.}}
  \bibinfo{volume}{6}, \bibinfo{number}{CSCW2}, Article
  \bibinfo{articleno}{448} (\bibinfo{date}{nov} \bibinfo{year}{2022}),
  \bibinfo{numpages}{21}~pages.
\newblock
\urldef\tempurl%
\url{https://doi.org/10.1145/3555549}
\showDOI{\tempurl}


\end{thebibliography}


\end{document}